%% file: paper.tex
\newif\ifndss
\ndssfalse

\newif\ifusenix
\usenixtrue

\newif\ifanonymous
\anonymousfalse

\ifndss
    \documentclass[conference]{IEEEtran}
    \usepackage[table,dvipsnames]{xcolor}
\fi

\ifusenix
    \documentclass[letterpaper,twocolumn,10pt,table,,dvipsnames]{article}
    \usepackage{usenix-2020-09}
    
    \usepackage{xcolor}
\fi

\newif\ifcomments
\commentstrue
\ifcomments

\else

\fi

\usepackage[pdf]{graphviz}
\usepackage{dot2texi}

\usepackage{wasysym}
\usepackage{graphicx}
\usepackage{subcaption}
\usepackage{listings}
\usepackage{color}
\usepackage{array,multirow,graphicx}
\usepackage{float}
\usepackage{numprint}
\npthousandsep{,}\npthousandthpartsep{}\npdecimalsign{.}
\definecolor{lightgray}{rgb}{.9,.9,.9}
\definecolor{darkgray}{rgb}{.4,.4,.4}
\definecolor{purple}{rgb}{0.65, 0.12, 0.82}
\definecolor{solablue}{HTML}{0068AD}
\definecolor{solaorange}{HTML}{FF5D00}
\definecolor{darkcandyapplered}{rgb}{0.64, 0.0, 0.0}

\usepackage{footnote}
\makesavenoteenv{tabular}
\makesavenoteenv{table}

\usepackage{hyperref}

\ifcomments
    \newcommand{\crs}[1]{[{\color{blue}Cris: #1}]}
    \newcommand{\agi}[1]{[{\color{green}Ági: #1}]}
    \newcommand{\saz}[1]{[{\color{red}Sazz: #1}]}
\else
    \newcommand{\crs}[1]{}
    \newcommand{\agi}[1]{}
    \newcommand{\saz}[1]{}
\fi
\newcommand{\nbPkgsFull}{\numprint{7605}}
\newcommand{\nbPkgsFalse}{\numprint{1173}}
\newcommand{\nbPkgsStudy}{  \numprint{6432}}
\newcommand{\nbPkgsCrossL}{\numprint{3494}}

\newcommand{\numCrashes}{\numprint{33}}
\newcommand{\nbVulnWapps}{six}

\newcommand{\numCWithSink}{\numprint{2802}}
\newcommand{\numCWithFlow}{\numprint{1669}}
\newcommand{\numCSanitizWithFlow}{\numprint{939}}
\newcommand{\numCVulnWithFlow}{\numprint{730}}

\newcommand{\numCrossLangAnalyzed}{\numprint{6401}}
\newcommand{\numCrossLangWithFlow}{\numprint{300}}
\newcommand{\numCrossLangVulns}{\numprint{111}}
\newcommand{\numCrossLangJSSanitiz}{\numprint{45}}
\newcommand{\numCrossLangCSanitiz}{\numprint{144}}
\newcommand{\numCrossLangBothSanitiz}{\numprint{22}}

\newcommand{\pkgsTimeoutC}{69}
\newcommand{\pkgsTimeoutJS}{34}

\newcommand{\nbCVEs}{seven}

\newcommand{\ratersKappa}{0.57}

\newcommand{\nbWinPkgs}{\numprint{1883}}
\newcommand{\pkgWithDirectExport}{\numprint{2956}}
\newcommand{\pkgDirectPerc}{45.83\%}

\lstdefinelanguage{JavaScript}{
	keywords={typeof, new, true, false, try, catch, function, return, null, catch, switch, var, if, in, while, do, else, case, break,let, const, throw},
	keywordstyle=\color{black}\bfseries,
	ndkeywords={class, export, boolean, throw, implements, import, this},
	ndkeywordstyle=\color{darkgray}\bfseries,
	identifierstyle=\color{Maroon},
	sensitive=false,
	comment=[l]{//},
	morecomment=[s]{/*}{*/},
	commentstyle=\color{darkgray}\ttfamily,
	stringstyle=\color{OliveGreen}\ttfamily,
	escapeinside={/*\#}{\#*/},	
	morestring=[b]',
	morestring=[b]",
	morestring=[b]`
}

\ifanonymous
    \author{}
\else
    \ifndss
        \author{\IEEEauthorblockN{Cristian-Alexandru Staicu}
        \IEEEauthorblockA{CISPA Helmholtz Center \\ for Information Security\\
        staicu@cispa.de}
        \and
        \IEEEauthorblockN{Sazzadur Rahaman}
        \IEEEauthorblockA{University of Arizona\\
        sazz@cs.arizona.edu}
        \and
        \IEEEauthorblockN{Ágnes Kiss}
        \IEEEauthorblockA{CISPA Helmholtz Center \\ for Information Security\\
        agnes.kiss@cispa.de}
        \and
        \IEEEauthorblockN{Michael Backes}
        \IEEEauthorblockA{CISPA Helmholtz Center \\ for Information Security\\
        backes@cispa.de}}
    \fi
    \author{{Cristian-Alexandru Staicu}\\
        CISPA Helmholtz Center \\ for Information Security\\
        staicu@cispa.de
        \and
        {Sazzadur Rahaman}\\
        University of Arizona\\
        sazz@cs.arizona.edu
        \and
        {Ágnes Kiss}\\
        CISPA Helmholtz Center \\ for Information Security\\
        agnes.kiss@cispa.de
        \and
        {Michael Backes}\\
        CISPA Helmholtz Center \\ for Information Security\\
        backes@cispa.de}
\fi

\newcommand{\toolname}{{\sc FlowJS}}

\newcommand{\mytikzmark}[1]{\tikz[remember picture] \node[coordinate] (#1) 
{#1};}
\usepackage{numprint}
\usepackage{tikz}
\usetikzlibrary{shapes.geometric}
\usetikzlibrary{arrows}
\usetikzlibrary{patterns}
\usetikzlibrary{shapes}
\usetikzlibrary{positioning}
\usetikzlibrary{tikzmark}
\usetikzlibrary{calc}
\usetikzlibrary[topaths]
\usetikzlibrary{arrows,backgrounds,snakes}
\usetikzlibrary{trees,positioning,shapes,shadows,arrows}
\usetikzlibrary{arrows.meta,positioning}

\definecolor{ggreen}{HTML}{9BBB59}
\definecolor{ppurple}{HTML}{9F4C7C}

\usepackage{pgfplots}
\usepackage{pgfplotstable}
\usepackage{filecontents}

\usepackage{dirtytalk}

\let\labelindent\relax

\usepackage{enumitem}

    \usepackage{booktabs}
    \usepackage{txfonts}

    \usepackage{mathtools}

\makeatletter

\long\def\ifnodedefined#1#2#3{%
    \@ifundefined{pgf@sh@ns@#1}{#3}{#2}%
}

\pgfplotsset{
    discontinuous/.style={
    scatter,
    scatter/@pre marker code/.code={
        \ifnodedefined{marker}{
            \pgfpointdiff{\pgfpointanchor{marker}{center}}%
             {\pgfpoint{0}{0}}%
             \ifdim\pgf@y>0pt
                \tikzset{options/.style={mark=*, fill=white}}
                \draw [densely dashed] (marker-|0,0) -- (0,0);
                \draw plot [mark=*] coordinates {(marker-|0,0)};
             \else
                \tikzset{options/.style={mark=none}}
             \fi
        }{
            \tikzset{options/.style={mark=none}}        
        }
        \coordinate (marker) at (0,0);
        \begin{scope}[options]
    },
    scatter/@post marker code/.code={\end{scope}}
    },
    box plot/.style={
        /pgfplots/.cd,
        black,
        only marks,
        mark=-,
        mark size=1em,
      	fill=solablue,
      	fill opacity=0.2,
        /pgfplots/error bars/.cd,
        y dir=plus,
        y explicit,
    },
    box plot box/.style={
        /pgfplots/error bars/draw error bar/.code 2 args={%
            \draw  ##1 -- ++(1em,0pt) |- ##2 -- ++(-1em,0pt) |- ##1 -- cycle;
        },
        /pgfplots/table/.cd,
        y index=2,
        y error expr={\thisrowno{3}-\thisrowno{2}},
        /pgfplots/box plot
    },
    box plot top whisker/.style={
        /pgfplots/error bars/draw error bar/.code 2 args={%
            \pgfkeysgetvalue{/pgfplots/error bars/error mark}%
            {\pgfplotserrorbarsmark}%
            \pgfkeysgetvalue{/pgfplots/error bars/error mark options}%
            {\pgfplotserrorbarsmarkopts}%
            \path ##1 -- ##2;
        },
        /pgfplots/table/.cd,
        y index=4,
        y error expr={\thisrowno{2}-\thisrowno{4}},
        /pgfplots/box plot
    },
    box plot bottom whisker/.style={
        /pgfplots/error bars/draw error bar/.code 2 args={%
            \pgfkeysgetvalue{/pgfplots/error bars/error mark}%
            {\pgfplotserrorbarsmark}%
            \pgfkeysgetvalue{/pgfplots/error bars/error mark options}%
            {\pgfplotserrorbarsmarkopts}%
            \path ##1 -- ##2;
        },
        /pgfplots/table/.cd,
        y index=5,
        y error expr={\thisrowno{3}-\thisrowno{5}},
        /pgfplots/box plot
    },
    box plot median/.style={
        /pgfplots/box plot
    },
}

\makeatother

\begin{document}

\title{Bilingual Problems: Studying the Security Risks Incurred by Native Extensions in \\ Scripting Languages} 

\maketitle

\begin{abstract}

Scripting languages are continuously gaining popularity due to their ease of use and the flourishing software ecosystems that surround them. These languages offer crash and memory safety by design, thus, developers do not need to understand and prevent low-level security issues like the ones plaguing the C code. However, scripting languages often allow \emph{native extensions}, which are a way for custom C/C++ code to be invoked directly from the high-level language. While this feature promises several benefits such as increased performance or the reuse of legacy code, it can also break the language's guarantees, e.g., crash-safety. 

In this work, we first provide a comparative analysis of the security risks of native extension APIs in three popular scripting languages. Additionally, we discuss a novel methodology for studying the misuse of the native extension API. We then perform an in-depth study of npm, an ecosystem which is most exposed to threats introduced by native extensions. We show that vulnerabilities in extensions can be exploited in their embedding library by producing reads of uninitialized memory, hard crashes or memory leaks in \numCrashes{} npm packages, simply by invoking their API with well-crafted inputs. Moreover, we identify \nbVulnWapps{} open-source web applications in which such exploits can be deployed remotely by a weak adversary.
Finally, we were assigned \nbCVEs{} security advisories for the work presented in this paper, most labeled as high severity. 

\end{abstract}
\sloppy


\input{introduction}

\input{threat_model}

\section{Misuse in Different Languages}
\label{sec:study}
To shed light on the pitfalls of existing native extension APIs, we build several simple extensions in three different scripting languages. These extensions are deliberately vulnerable, attempting to stress the corner-cases of the API, e.g., by omitting type checks on values coming from the scripting language. We then attempt to break the safety of the scripting language by providing well-crafted values to the vulnerable extension's methods, i.e., we assume a strong attacker model in this section. Finally, we observe whether the API actively tries to prevent the exploitation and if so, in which way. For creating the list of misuses, we draw inspiration from the work of Brown et al.~\cite{BrownNWEJS17} for JavaScript bindings, but we also add several misuses that are specific to native extensions, e.g., read-write local variables. For the study, we use Node.js 15.4.0, Python 3.8.5 and Ruby 2.7.0p0. For Node.js we consider two different native extension APIs, i.e., Nan\footnote{\url{https://www.npmjs.com/package/nan}} and N-API\footnote{\url{https://nodejs.org/api/n-api.html}}, due to their prevalence in open-source projects.

In Table~\ref{tab:comparative-analysis}, we provide an overview of our findings. 
We mark each misuse type with a unique identifier ($M_i$) and will use these throughout the paper referring to them.
One can see that there is a lot of variation among the considered languages, i.e., while some prevent most of the misuses by construction, others put the burden of using the API in a safe way on the developer. Nevertheless, none of the languages prevents all misuses. For example, a crash in the native extension compromises the availability of the application relying on it in all considered APIs. 
Below, we discuss in detail each misuse class and how they are handled by different APIs.

\newcommand*\emptycirc[1][1ex]{\tikz\draw (0,0) circle (#1);} 
\newcommand*\halfcirc[1][1ex]{%
  \begin{tikzpicture}
  \draw[fill] (0,0)-- (90:#1) arc (90:270:#1) -- cycle ;
  \draw (0,0) circle (#1);
  \end{tikzpicture}}
\newcommand*\fullcirc[1][1ex]{\tikz\fill (0,0) circle (#1);} 

\begin{table*}[t]
	\center
	\small
	\caption{
		Different misuses of the native extension API and their prevalence in the considered languages. \fullcirc \, means that the API allows the misuse, \halfcirc \, that the API partially allows it, and \emptycirc \, that the API prevents the misuse. We estimate the severity based on the impact a given misuse might have on the security, privacy, or availability of a web application.
	}
	\begin{tabular}{lll ccccc}
		\toprule
		Type &Id & Misuse  &   Node.js-N-API &   Node.js-Nan   & Python & Ruby & Severity \\
		\midrule
		\parbox[t]{2mm}{\multirow{2}{*}{\rotatebox[origin=c]{90}{Errors}}} & $M_1$ & Not catching C++ exceptions & \fullcirc & \fullcirc & \fullcirc & N/a & Low\\
		& $M_2$ & Not handling runtime errors in C/C++ & \fullcirc  & \fullcirc & \fullcirc & \fullcirc & Medium\\
		\midrule
		
		\parbox[t]{1cm}{\multirow{4}{*}{\rotatebox[origin=c]{90}{Arguments}}} & $M_3$ & Passing arguments with a wrong type & \fullcirc & \fullcirc & \emptycirc & \emptycirc & High\\
		& $M_4$ & Passing wrong number of arguments & \fullcirc & \halfcirc & \emptycirc & \emptycirc & High\\
		& $M_5$ & Not accounting for different semantics of \textbackslash0 & \fullcirc & \fullcirc & \emptycirc & \emptycirc & High\\
		& $M_6$ & Passing arguments that overflow numeric types & \fullcirc & \fullcirc & \emptycirc  & \emptycirc & High\\
		\midrule
		
		\parbox[t]{2mm}{\multirow{2}{*}{\rotatebox[origin=c]{90}{Ret.}}}& $M_7$ & Missing return statement & \fullcirc & \emptycirc & \halfcirc & \fullcirc & Low\\
		& $M_8$ & Declaring interface methods that return void & \emptycirc & \emptycirc & \halfcirc & \emptycirc & Low\\
		
		\midrule
		\parbox[t]{2mm}{\multirow{2}{*}{\rotatebox[origin=c]{90}{Mem.}}}& $M_9$ & Returning uninitialized memory values & \fullcirc & \emptycirc & \halfcirc & \emptycirc & Medium\\ 
		& $M_{10}$ & Mismanagement of cross-language pointers & \fullcirc  & \emptycirc & \fullcirc & \emptycirc & Low\\		
		
		\midrule
		\parbox[t]{2mm}{\multirow{2}{*}{\rotatebox[origin=c]{90}{\parbox{0.7cm}{\centering High-level}}}}& $M_{11}$ & Producing unexpected side-effects in the runtime & \emptycirc & \emptycirc & \emptycirc & \fullcirc & High\\
		& $M_{12}$ & Blocking the runtime with slow cross-language calls & \fullcirc & \fullcirc & \fullcirc & \fullcirc & Medium\\
		\midrule		
		
		\parbox[t]{2mm}{\multirow{5}{*}{\rotatebox[origin=c]{90}{\parbox{1.4cm}{\centering Low-level}}}} & $M_{13}$ & Reading outside of an allocated buffer & \fullcirc & \emptycirc & \emptycirc & \emptycirc & High\\ 
		& $M_{14}$ &Using a pointer after it was freed & \fullcirc & \fullcirc & \fullcirc & \halfcirc & High\\ 
		& $M_{15}$ &Freeing a pointer twice & \halfcirc & \halfcirc & \halfcirc & \halfcirc & High\\ 
		& $M_{16}$ &Failing to deallocate unused memory & \fullcirc & \fullcirc & \fullcirc & \fullcirc & Low\\
		& $M_{17}$ &Interpreting user input as format string & \halfcirc & \halfcirc & \emptycirc & \emptycirc & High\\
		\bottomrule
	\end{tabular}
	\label{tab:comparative-analysis}
\end{table*}

\textbf{Error containment.} As mentioned earlier, scripting languages follow a no crash philosophy. For example, in case of division by zero, Ruby and Python produce an exception that can be gracefully handled in a \texttt{try-catch} block, while JavaScript simply outputs the \texttt{Infinity} value. Moreover, in Node.js, developers often rely on a process-level exception handler that prevents any unexpected exception from crashing the application. We believe that this crash avoidance mentality has to do with the main use case of scripting languages, i.e., writing web applications, for which availability is one of the most important requirements. Native extensions can violate this no crash philosophy in two ways: by producing low-level crashes ($M_2$) that terminate the whole process or by leaking low-level exceptions ($M_1$) that cannot be handled by a \texttt{try-catch} block in the scripting language. Let us consider the \texttt{int64-napi} npm package that wraps the \texttt{int64} C type. It provides a \texttt{divide} method that can be invoked as follows:
\begin{lstlisting}[numbers=none]
const int64 = require('int64-napi');
const Int64 = int64.Int64; 
try {
 int64.divide(10, 0); // hard crash of Node.js
} catch(e) { } // never invoked
\end{lstlisting}
This code snippet produces a hard crash that cannot be handled in the corresponding \texttt{try-catch}. Such an outcome may surprise users that consider a catch clause as a universal safety net. Similarly, if there are C++ exceptions that are not properly handled by the native extension, there is no way for the scripting language to catch them. We saw this behavior in all the languages that support C++ native extensions. 

\textbf{Argument translation.} Since the analyzed scripting languages are weakly-, dynamically-typed, while C/C++ is strongly-, statically-typed, the native extension API has to assist the user in translating between these two type-systems. In Ruby, one needs to specify the number of arguments at extension declaration time, while in Python, the API for retrieving the arguments mandates that the user specifies the number of arguments ($M_4$) and their type ($M_3$). Any violation of these specifications would result in aborting the current method invocation. By contrast, in Node.js, both considered APIs specify that the users should voluntarily check the arguments' types and their number, and decide when to proceed. As seen in Figure~\ref{fig:example}, this may lead to serious problems such as processing strings with negative length or even worse, user-provided values considered as object pointers. We direct the reader to
~\cite{BrownNWEJS17} for an extensive discussion about the implications of breaking type safety in V8-based runtimes. 
We further stress the fundamental differences in the way errors are signaled in the different scripting languages. Whenever a mismatch is detected between the requested type for a value and its dynamic type, Python and Ruby stop immediately with an exception. One of the two considered native extension APIs for Node.js, N-API, returns a non-empty status code, while the other, Nan, does not detect the mismatch.
Even when the types are correctly aligned, there are still problems caused by the different ways in which a given type is represented in the two languages. While the null terminator \texttt{{\textbackslash}0} can appear in valid strings of the considered scripting languages, in C it marks the end of a string. Hence, if such characters are allowed to freely cross the language boundary ($M_5$), as it is the case in Node.js, they may allow attackers to strip important information from a value or to cause confusion about the string length as illustrated in Figure~\ref{fig:example}. Ruby and Python refuse to continue with the invocation when such characters are detected. A similar issue appears when a numeric value  overflows ($M_6$) due to a mismatch in the types' capacity. This case is prevented again by Ruby and Python, but allowed in Node.js. Integer overflow may invalidate important checks performed in the scripting language, e.g., \texttt{val>0}, since the invariant may not hold anymore for the translated value.

\textbf{Missing return.} 
To our surprise, there are also subtle bugs involving the return value of a function. A missing return statement~($M_7$) causes a hard crash when reading the return value in Python, Ruby and N-API. This may surprise developers who expose the native extension directly to their clients and never test for such corner cases. Returning null values from the extension does not cause problems in the analyzed languages, but declaring the return value as \texttt{void} ($M_8$) causes a hard crash on method invocation in Python.  

\textbf{Memory issues.} Similarly to the example in Figure~\ref{fig:example}, native extensions may expose non-initialized memory areas to the scripting language ($M_9$). Such memory locations may contain sensitive user information available in the process. In N-API, one can expose both uninitialized string values and buffers, while in Python only buffers are allowed. Ruby and Nan proactively initialize such memory areas with null bytes. Memory issues may also appear due to the garbage collector not freeing pointers to interface objects, exchanged across the language boundary ($M_{10}$). For primitives, all considered APIs prevent this by default. Python, however, makes it easy to overwrite this behavior by claiming ownership of certain pointers. While this is not a problem by itself, carelessly using this feature may compromise the availability of the entire application. Another interesting case is classes being exposed from C/C++ to the scripting language. When using Node.js (N-API), the garbage collector {does not free references} that are declared using \texttt{ Napi::Persistent} API. As discussed further in Section~\ref{subsec:exploits}, we identify several real-world libraries that are misusing this API, causing memory leaks.

\textbf{High-level issues.} Most of the considered APIs expose only opaque pointers to the C/C++ world. That is, the native extensions cannot directly access the exact memory location of an object, nor can they modify it without the aid of the API. In Ruby, however, one can obtain a raw pointer that allows the modification not only of the argument passed to the extension, but also of other variables defined in the same memory region ($M_{11}$). In this way, a problematic extension may access or even alter encapsulated values. 
Considering that many developers use native extensions for heavy computation, e.g., cryptographic operations,
it is somewhat surprising that the default behavior of all the considered APIs is to invoke the extension in a synchronous manner ($M_{12}$). That is, the main thread of the scripting language is blocked until the native extension computes. This may lead to serious availability issues if an attacker can control the amount of work the extension performs, especially in Node.js.

\textbf{Low-level issues.} Finally, we consider a handful of low-level vulnerabilities in our study to see if different APIs hinder their exploitation or not. To our dismay, in N-API, we could exploit a textbook buffer overflow ($M_{13}$) to overwrite local variables defined in the native extension. We also note that use-after-free ($M_{14}$) is allowed in most of the languages, but Ruby seems to initialize the freed memory areas with null bytes (we observe a similar behavior in case of uninitialized memory (${M_9}$)). 
A double free ($M_{15}$) always triggers a core dump, and a format string vulnerability ($M_{17}$) is usually prevented by the compiler. However, in Node.js, only a warning is produced, while in other languages the compilation is aborted. Finally, none of the APIs make any effort to prevent or detect memory leaks in the extension code itself ($M_{16}$).

\input{figs/overview}

\textbf{Summary.} 
As an artifact of the presented study, we provide a set of benchmarks as supplementary material\footnote{\url{https://native-extension-risks.herokuapp.com/}}, exemplifying each misuse in a separate native extension, for all considered scripting languages. We believe that this suite can be useful both for users trying to understand the pitfalls of each API, and for language designers to inform their design decisions.

Considering the presented findings, we conclude that there is a lot of variation in the implementation of native extensions in various languages. Some of the APIs put a lot of effort in preventing users from misusing them, while others are more permissive. Node.js, in particular, seems to be very liberal in its API's design, outsourcing most of the safety checks to the developers. We filed a security issue summarizing our findings to the Node.js developers, who appreciated our report as informative, they argue that the identified issues are not security problems of the API, but of the packages misusing it. As we show in Section~\ref{subsec:exploits}, this relaxed design decision is the main enabler for several security issues in popular npm packages.
The Node.js maintainers promised, however, to fix some of the identified issues, e.g., the behavior responsible for $M_6$.
While the misuses presented in this section aim to emulate realistic user interactions, the reader may wonder whether such cases appear in practice and if so, if they affect real-world applications.  We now proceed to designing a methodology for studying this aspect.

\input{methodology}

\section{Evaluation}
\label{sec:evaluation}
To study the security impact of native extension misuse and the feasibility of its automatic detection, we first present an empirical study of missing type checks ($M_3$, $M_4$) in npm packages  (Section~\ref{subsec:npm-study}). We then identify multiple zero-day vulnerabilities in real-world packages and report them to their maintainers (Section~\ref{subsec:exploits}). Finally, we show that vulnerabilities in libraries can be exploited remotely in web applications (Section~\ref{subsec:webapps}).

\subsection{Missing type checks in npm}
\label{subsec:npm-study}
{As discussed in Section~\ref{sec:pkg-analysis}, we study the feasibility of automatic misuse detection by focusing on an important class of misuses: missing type checks ($M_{3}$, $M_{4}$). This allows us to draw relevant conclusions about our hypothesis, without investing tremendous engineering effort into modelling the APIs corresponding to all the misuses in Table~\ref{tab:comparative-analysis}.}

\paragraph{Setup.}To identify packages that are likely to contain native extensions, we analyze the entire npm graph available on 9\textsuperscript{th} of February 2021.
We consider all the packages that directly depend on five popular helper packages that are widely used for developing native extensions: \texttt{bindings}, \texttt{node-gyp}, \texttt{prebuild-install}, \texttt{node-addon-api} and \texttt{nan}. While this approach may have some false negatives, it is a cost-effective way to identify packages of interest without the need to download {and analyze} all the 1.5 million packages on npm and instead, allocate more resources for the in-depth study. In total, we download \nbPkgsFull{} npm packages that comply with the aforementioned  requirement.
Of these, we identify \nbPkgsFalse{} false positives that do not contain any C/C++ code. 
After excluding these packages, we are left with \nbPkgsStudy{} packages that we further use in our study. We believe that this is a large enough sample for drawing conclusions about how native extension APIs are used in Node.js. For each package, we download and analyze its latest version at the time of our study. In Appendix~\ref{ap:perf} we show the performance of graph extraction for the two languages (JavaScript and C/C++).


To investigate the effect of missing type checks ($M_3$), we first search in all the C++ files for calls to type conversion APIs. We note that there are multiple ways in which arguments coming from JavaScript can be converted to a given type, e.g., \texttt{*.As<Type>}, \texttt{*.To<Type>}, and not all of these APIs react in the same way to misuse, but they all proceed with an unsafe value. In Figure~\ref{fig:types_package} in Appendix~\ref{app:numpack}, we depict the total number of packages that explicitly convert values to a given type.
%
We then perform intra-procedural analysis on the C/C++ native extensions, namely on the output \texttt{.dot} graphs of Joern, for detecting missing type checks.
In total, we identify \numCWithSink{} packages with type conversions, of which \numCWithFlow{} have a flow to the conversion API. Of these, \numCSanitizWithFlow{} were type checked in the native extension code, and \numCVulnWithFlow{} were not. 

To evaluate our tool's effectiveness at finding flows to relevant APIs, we perform a controlled experiment with single C++ functions, all containing a flow to the target APIs. We collect a set of 25 functions from real-world packages, aiming for a diverse set of code constructs, e.g., macros, different APIs, intermediary variables, or chained calls. We then run our tool on these benchmarks and detect a flow in 21 of them (84\% recall). One failure was due to Joern's imprecise data flow graph that fails to capture an important data flow step, one due to a missing sink, and two due to our analysis' inability to handle type checks or sinks present in macros or functions. We provide the set of microbenchmarks in the supplementary material to assist replication.
We also verify all the produced flows and find a single false positive caused by lack of path sensitivity (95\% precision).
We conclude that our prototype can handle a wide range of code constructs present in real-world code, producing a manageable amount of both false positives and negatives. We next proceed to identifying zero-day vulnerabilities.


\input{figs/type_conv}

To this end, we concentrate on three APIs that produce hard crashes on misuse. Since we want to manually validate each finding,
it is easier to judge the presence of a crash than the success of other types of payloads, e.g., the effect of integer overflow. We note that most work in the fuzzing domain uses similar testing oracles. 
In Figure~\ref{fig:typecheck}, we show the total number of npm packages that contain detected data flows to (i) \texttt{*.ToLocalChecked()}, which is a method on the V8's \texttt{Maybe} type that concretizes a given value, (ii) APIs for casting to Buffer and (iii) APIs for casting to function.
We depict both sanitized (grey bar on the right) and unsanitized flows (colorful bar on the left) and further categorize unsanitized flows based on their exploitability after manual verification. Appendix~\ref{ap:manual} details the exact steps of our manual analysis.
Figure~\ref{fig:typecheck} shows first of all that 63\% of the packages with a flow to type conversion APIs, also type check the arguments. Out of these, 82\% do so in C/C++, showing that indeed checking often occurs in the proximity of the API usage. 
These findings are good news since they show that most developers are aware of this best practice of the API. We successfully exploit 38 flagged packages, showing the feasibility of our automated approach. We identify a total of 22 clear false positives, all of them caused by type checks placed in different functions / macros. This implies a false positives rate of 6\%, in line with our controlled experiment. It also suggests that our lightweight analysis design (path-insensitive, intra-procedural) is adequate for the problem at hand. However, these results should be taken with a grain of salt since for the majority of the reports we could not verify their exploitability empirically and thus, we rely on manual code analysis to judge whether the report is a false positive or not. 



The main benefit of performing cross-language analysis, as described in Section~\ref{sec:pkg-analysis}, is to automate the first part of our manual process: the analysis should assign all the packages assigned to \say{type check in JavaScript} in Figure~\ref{fig:typecheck} (depicted in blue) to the grey bar. Another more subtle benefit is the improvement in user experience for the analyst. We found ourselves often switching  between the C/C++ file and the JavaScript part of a package during manual analysis. A cross-language visual representation of the code would significantly ease this process.
Using our prototype, we analyzed \numCrossLangAnalyzed{} cross-language flows, and detected \numCrossLangWithFlow{} flows to the sink. 
Out of these, \numCrossLangCSanitiz{} sanitize in C/C++, \numCrossLangJSSanitiz{} sanitize in JavaScript, \numCrossLangBothSanitiz{} sanitize in both, and \numCrossLangVulns{} do not perform any sanitization.
We provide the obtained cross-language data-flow graphs in the supplementary material of this paper\footnote{\url{https://native-extension-risks.herokuapp.com/}} to increase confidence in our analysis method. Furthermore, we show graphs for three examples in Appendix~\ref{sec:app}: an unsanitized flow, a flow sanitized in JavaScript, and one sanitized in both languages.


\begin{table}
\setlength{\tabcolsep}{3pt}
    \caption{The most important npm packages for which we identified a zero-day vulnerability.
    Vulnerabilities \textcolor{red}{$M_2$} and \textcolor{red}{$M_9$} were found through manual inspection of flagged packages.
    }
    \label{fig:imp-exploits}
  \rowcolors{2}{gray!25}{white}
    \centering
    \resizebox{1\columnwidth}{!}{%
    \begin{tabular}{c|c|c|c|c|c}
         \rowcolor{gray!40}
         \textbf{Package Name} & \textbf{\#Downloads\footnote{Weekly downloads at the time of writing}} & \textbf{Misuse} & \begin{tabular}{c}
              \textbf{Remotely} \\
              \textbf{Exploitable?}
         \end{tabular}  & \textbf{Status}\\
         \hline
bignum	& 5,091 & $M_3$ & Yes & CVE-2022-25324\\
ced & 1,765 & $M_3$ & No & CVE-2021-39131\\
libxmljs & 28,629 & $M_3$ & Yes & CVE-2022-21144\\
sqlite3	& 452,737 & $M_3,$\textcolor{red}{$M_9$} & Yes & CVE-2022-21227\\
pg-native & 92,436 & $M_3$ & Yes & CVE-2022-25852\\
utf-8-validate & 917,251 & $M_3,M_4$ & No & Reported\\
@discordjs/opus & 63,007 & \textcolor{red}{$M_2$}$,M_3,$\textcolor{red}{$M_9$} & Yes & CVE-2022-25345\\
fast-string-search & 25 & $M_3$ & Yes & CVE-2022-22138\\
time & 1,701 & $M_3$ & Yes & Reported\\
bigint-buffer & 159,067 & $M_3$ & Yes & Reported\\
\hline
    \end{tabular}
    }
\end{table}

\subsection{Zero-day vulnerabilities}
\label{subsec:exploits}

We identify a total of \numCrashes{} libraries for which we can exploit a misuse through their public API, i.e., under the strong attacker model. We invite the reader to consult the complete list of affected packages in Appendix~\ref{ap:list-all}, while in Table~\ref{fig:imp-exploits} we show our most important findings. In all the \numCrashes{} libraries we first detect an issue caused by careless type conversions ($M_3$, $M_4$), and for three of them we identify additional types of misuses ($M_2$, $M_9$), using the supplementary manual analysis described in Section~\ref{sec:pkg-analysis}. To confirm that all the identified vulnerabilities caused by misuse are indeed security-relevant and worth-fixing, we approached the maintainers of the library to report our findings. We describe below the disclosure process and the outcome of these interactions.

\paragraph{Vulnerability disclosure.}
We have reported the discovered vulnerabilities to the maintainers using their email address provided in the package description. We have received a few responses and a CVE with this strategy, and after the grace period of six months have passed, we worked together with Snyk\footnote{\url{snyk.io}} to report the vulnerabilities affecting the most high-profile packages depicted in Table~\ref{fig:imp-exploits}. This disclosure strategy has resulted in an additional six~CVEs. All CVE-assigned high-profile packages have also been fixed thanks to our disclosure efforts.
The maintainer of \texttt{utf-8-validate} dismissed our report saying that it is unlikely that an attacker can trigger this bug remotely, while for the other two we reported we are still pending responses. Six CVEs were assigned a high severity label and one of them medium. These interactions show that developers acknowledge the risks posed by native extensions and are willing to mitigate it, when provided with actionable reports.
%
Below we discuss three examples illustrating security-relevant consequences that can be obtained by exploiting the identified native extension misuses.

\paragraph{Uninitialized memory.} Let us consider discuss a type confusion vulnerability with surprising security implications. \texttt{fast-string-search} is a package that promises to be ``10 times faster than the \texttt{indexOf} function of a Node.js string''. Using our automated approach, we detect missing type checks corresponding to the two arguments of the \texttt{indexOf} method. The sequence of API calls for retrieving the string from JavaScript is very similar to the one in Figure~\ref{fig:example}.
Thus, if an attacker passes numbers instead of strings, a very large string length is retrieved in the first native extension call, which will then crash the process when trying to allocate a C string that large. After investigating further, we notice that the compiler reuses memory locations between API calls and hence, strings can be leaked between calls:
\begin{lstlisting}[numbers=none]
const fss = require('fast-string-search');
fss.indexOf("My password is Foo123#", "is");
let res = fss.indexOf(1, "Foo123"); 
console.log(res); // prints 15
\end{lstlisting}
Let us assume the call in line 2 is performed using user's arguments, while the one in line 3 uses arguments under attacker's control. The first call correctly returns 12, the position of the substring "is" in the larger string containing the password. The second call, though, passes an invalid string as a first argument and a password guess as second argument. In this case, the libraries reuse the argument provided in the first call, thus the result of the second call is 15 - the position of "Foo123" (passed in the second call) in the string "My password is Foo123\#" (passed in the first call). Therefore, the attacker can guess parts of the user's password in this way. In general, when prevented with invalid arguments, this npm package reuses uninitialized memory as input. Exploiting such bugs in real systems can have a detrimental impact on user's privacy, similar to that of vulnerabilities caused by the deprecated \texttt{Buffer} constructor\footnote{\url{https://snyk.io/blog/exploiting-buffer/}}.
We believe this is an extremely dangerous behavior for a third-party package in a scripting language. Most users of such packages are not familiar with such subtle memory-related bugs, and would not expect to see them surface in their scripting language. We stress once again that developers can import native extensions carelessly with \texttt{npm install fast-string-search}, most of them probably not knowing that they use native code under the hood. We also note that this bug is not possible in any of the other scripting languages we analyzed in Section~\ref{sec:study}, and it is a consequence of the permissive API design of Node.js.

\paragraph{Hard crash.}
Let us consider the following proof of concept using the modular exponentiation in the \texttt{bignum} package:
\begin{lstlisting}
let bignum = require('bignum');
try {    
    bignum(10).powm(1, {});
} catch(e) { 
    console.log(e); // never executed
}
\end{lstlisting}
When this API is provided with an object literal instead of a number, it instantly crashes the Node.js process, without the possibility of recovering, e.g., error handling. 
Prior work reports that Node.js' parallelism relies on a single point of failure, i.e., the event loop~\cite{DavisWL18, StaicuP18}. 
In the context of web applications, the impact of a hard crash on the server-side is an instant halt of the server. This implies dropping all pending requests, and hence loss of precious data. 
Restarting the server may take several seconds in which user requests cannot be served, seriously impacting the availability of the server. If, however, the server runs as part of a cloud platform with autoscaling, a new server instance needs to be spawned immediately and the current one has to be restarted. By sending multiple such requests, an attacker can mount a Yo-Yo attack\cite{Bremler-BarrBS17}, causing significant economic damage.

\paragraph{Memory leak.}
Let us consider the following example, showing a memory leak in one of the analyzed packages:
\begin{lstlisting}
const { OpusEncoder } = require('@discordjs/opus')
while (1) new OpusEncoder(48000,3); 
\end{lstlisting}
When monitoring the memory consumption for this simple code snippet, we observe a linear growth, indicating the lack of garbage collection. After few seconds, the machine becomes extremely slow and then the process crashes. While this example is artificial, if this package is used as part of a web application and such objects are created in response to user requests, the attacker can mount a memory-based DoS attack. This would in turn slow down or even stop the event loop like in the case above, causing a hard-to-diagnose performance problem that would impair Node.js' parallelism.
The root cause of the problem is the ObjectWrap API\footnote{\url{https://nodejs.org/api/n-api.html\#object-wrap}} in Node.js (N-API), in particular the \texttt{Napi::Persistent} method, which impairs the garbage collector. Therefore, fixing the indicating problem requires migrating away from that API or from all the packages using it. We found a similar problem in the \texttt{sqlite3} package.

\subsection{Impact on web applications}\label{subsec:webapps}

Vulnerabilities in native Node.js extensions motivated us to measure the problem's impact on web applications' security. In this section, first, we present our application selection criteria and pre-processing steps. Then, we discuss our findings. Note that, given a web application using a vulnerable extension,~\toolname{} can be used to detect if the vulnerability can be exploited remotely. However, since, it requires the prior knowledge about the vulnerability, we only run experiments on the ones that are manually confirmed.

\textbf{Application selection and pre-processing.}
For this experiment, we select seven most used (> 1,000 downloads) vulnerable extensions from Table~\ref{fig:imp-exploits} that can be exploited remotely: \texttt{sqlite3}, \texttt{libxml}, \texttt{bignum}, \texttt{time}, \texttt{pg-native}, \texttt{discord/opus} and \texttt{bigint-buffer}. Next, our goal is to measure their impact on web applications that are using them. For each extension, we retrieve their dependent applications from GitHub. 
We download at most 300 repositories per vulnerable package, consisting of in total 1,993 Node.js applications (Table~\ref{vulnerable:apps}). 
Note that, a dependent repository does not imply a web application. However, sorting web applications from others would require manual effort, which we conservatively employ -- if and only if \toolname{} reports an alert. 

Our current prototype, \toolname{}, can only analyze one JavaScript file at a time. To find vulnerabilities across multiple file, \toolname{} requires the files to be merged. We use Google Closure Compiler to merge all the JavaScript files from a repository before running \toolname{}. However, setting up a repository properly for Closure, would require properly setting up the custom module dependencies, proper handling of duplicate variable declarations, etc, which requires non-trivial manual efforts. \textit{Without such efforts, Closure successfully merged 1,141 out of 1,993 repositories.} If Closure fails to merge files for a repository, we analyze each of the files separately. Note that, this is a fundamental limitation of single-pass compiler platforms like Closure that we used to implement \toolname{}. They work on the boundary of one translation unit (TU)---at the file level and miss cross-TU flows. Solving this issue would require designing a multi-pass analysis of our own, which comes with its own challenges (similar to merging files by Closure). 

{
\textbf{Exploitable misuses as program flows.} A common property of all the selected native extension APIs is that the vulnerability can be triggered if an attacker can control the input to them. In web applications, an attacker can control the request data. If unsanitized request data is passed to those APIs, then an attacker can turn the vulnerabilities into exploits. Based on this insight, we use \toolname{} to find unsanitized flows from request data to the vulnerable APIs. We report a misuse if an element of the network request directly influences the API parameter of interest. In Table~\ref{vulnerable:apps}, we provide the APIs corresponding to each rule specification.}

\begin{table}[]
\caption{Vulnerable applications per package. TP refers to true positives.}\label{vulnerable:apps}
\resizebox{\columnwidth}{!}{%
\begin{tabular}{@{}lllll@{}}
\toprule
Package  & Criteria (Sinks)               & \#Repos & \#Misuses & \#Exploitable  \\ 
  &                &  & (Total/TP) & (Total/TP) \\ \midrule
\texttt{sqlite3} & \texttt{run(\_, \textbf{data})} &  283    &   4/3                &   4/3                \\
\texttt{libxml}   & \texttt{parseXml(\textbf{xml})} & 296     &    2/2               & 2/2                   \\
\texttt{bignum}   & \texttt{powm(\_, \textbf{pow})}          &  293  &           0/0        &      0/0             \\
\texttt{time}     & \texttt{setTimezone(\textbf{tz})} & 298  & 0/0 & 0/0 \\
\texttt{pg-native}     & \texttt{query(\_, \textbf{values}, \_)} & 270  & 1/1 & 1/1 \\
\texttt{discordjs/opus}     & \texttt{encode(\textbf{data})} & 272  & 0/0 & 0/0 \\
\texttt{bigint-buffer}     & \texttt{toBigIntLE(\textbf{buff})} & 282  & 0/0 & 0/0 \\
\bottomrule
Total    & -- & 1,993  & 7/6 & 7/6 \\
\bottomrule
\end{tabular}
}
\vspace{-15pt}
\end{table}

\textbf{Our findings.} 
To find misuses corresponding to each of the selected APIs, we create the corresponding rule specification. Then, we run \toolname{} with the rule specifications on the selected GitHub repositories. Table~\ref{vulnerable:apps} presents the summary of our experimental findings. 
\toolname{} reported four exploitable vulnerabilities in four applications (out of 283) using \texttt{sqlite3}, two vulnerabilities in two applications using \texttt{libxml} and one vulnerability in one application using \texttt{pg-native}. \toolname{} did not find any vulnerabilities in any other categories. Our manual investigation shows that six out of seven alerts in six applications are true positives (Table~\ref{vulnerable:apps}). In the false positive case also, data from the request attributes are directly passed to the \texttt{sqlite3} API, however, before doing so, a type check is performed. {Since our current implementation of \toolname{} is not path-sensitive, it cannot detect such type-checking constraints. Therefore, it raised an alert.} We show the source code of the false positive and additional details in Appendix~\ref{sec:app2}. {We also ran \toolname{} to find how often \texttt{libxml} is used on content read from local files.}
Our analysis found that 27 applications directly read such files, which may be security-relevant for some web applications. 

Below we provide an example misuse of \texttt{sqlite3}'s \texttt{run} API, detected by our approach. This code is protected against SQL injection by the use of prepared statements. However, the vulnerability in  \texttt{sqlite3} allows attackers to trigger hard crashes remotely, e.g., by providing the value \texttt{\{toString: 23\}} for the \texttt{img} attribute of the request's body.

\begin{lstlisting}
server.post("/", (req, res) => {
  const {img,title,category,description,link} = req.body
  const query = `INSERT INTO ideas (image, title, category, description, link) VALUES (?,?,?,?,?)`
  const values = [img,title,category,description,link]
  db.run(query, values, function(err) {
   if(err) {
      console.log(err)
      return res.send("Erro no banco de dados")
    }
    return res.redirect("/ideias")
  }) ...
})
\end{lstlisting}
\vspace{-10pt}

Our results show that after identifying an exploitable misuse in an npm package, an adversary can further detect vulnerable endpoints of open-source web applications that rely on this package. To prevent such exploitation, it is crucial for the community to detect and patch vulnerable npm packages.

\section{Discussion}
\label{sec:discussion}


\textbf{Impact of security findings.} Our success with identifying high-severity vulnerabilities in native extensions of popular libraries imply that even well-maintained libraries struggle with using native extensions securely. The empirical evidence clearly shows that most bugs are caused by the permissive nature of Node.js. Hence, we recommend that maintainers of this runtime reassess whether this design is in the best interest of their users.  

\textbf{Feasibility of the weak attacker model.} Prior work ~\cite{dinh2021favocado,BrownNWEJS17,ParkDGNVF20,HollerHZ12,KocherHFGGHHLM019} assumes a strong adversary that can run arbitrary high-level code to exploit bugs in the native layer. To the best of our knowledge, we are the first to propose a methodology for finding low-level vulnerabilities that can be exploited remotely by weak, web attackers.
By showing the feasibility of this scenario, our findings raise concerns about a hidden attack surface of web applications written in scripting languages.
Code analysis tools should challenge the assumption that low-level code is trustworthy. Additionally, the community should continuously vet packages with native extensions for both vulnerable and malicious code.

\textbf{Extensibility of the prototype.} {The developed prototype} is by no means a complete solution for identifying security problems caused by native extensions in scripting languages. There are several components that can be improved by future work. We opted in our design for a path-insensitive, intra-procedural analysis. While very scalable, this strategy results in a non-negligible number of false positives, which can be reduced by employing more sophisticated analysis techniques.
Also, while this simple analysis may suffice for identifying missing type checks, it is not enough for detecting more complex problems, such as use-after-free or buffer overflow. {That is because type conversions often appear at the language boundary, while unsafe buffer operations may appear anywhere in the program.} While we provide initial evidence that cross-language analysis can aid analysts in the vulnerability detection task, we recognize that in its current form, this method might be costly for practitioners. A taint summarization  approach~\cite{ArztB16,StaicuTSMP20} might scale better. Similar limitations exist in~\toolname{} too. This is because to run large-scale analysis we traded precision and soundness for scalability. For example,~\toolname{} is flow- and path-insensitive, which hurts its precision. Additionally, incorporating alias analysis in~\toolname{}  would result in better soundness.


\textbf{Applying methodology to other languages.} We believe that our high-level methodology that emphasises the analysis of libraries can be applied to any other scripting language. 
To add support in our prototype for other scripting languages, one can reuse the C/C++ extraction and the post-processing of the dot graphs, i.e., the graph traversals. However, for each scripting language one would additionally need: (i) a data flow extraction tool that can produce graphs in the dot format, (ii) a way to identify the program locations in which the native extensions are invoked, so that the cross-language graphs can be generated, and (iii) additional modelling to identify security-relevant sinks and sanitizers. While this can be done with sufficient engineering effort for all the languages studied in Section~\ref{sec:study}, we believe that our results for npm suffice for drawing conclusions about the feasibility of the methodology.

\input{related-work}

\section{Conclusions}
\label{sec:conclusions}

In this work, we first systematically analyze the pitfalls of using native extensions in three scripting languages. We show how a failure to adhere to best practices can cause serious problems such as an exploitable buffer overflows or modifications of encapsulated values of the scripting language. 
We further propose a methodology to systematically detect misuses of a native extension API. By leveraging that, we show how the security problems propagate in the dependency chain, first to the enclosing library, and  then to the web application relying on it. 
We show that many libraries fail to type check arguments coming from the scripting language and that attackers can cause uninitialized memory reads, hard crashes or memory leaks by providing well-crafted inputs to the library API. In total, we create proof-of-concept exploits in \numCrashes{} real-world npm packages and show that some of the vulnerabilities could be exploited remotely in  open-source web applications.
This paper is first of all a warning for developers: native extensions in third-party code may violate all your assumptions about the safety of the scripting language you use. To put it more poetically, \emph{tell me what you include, so I can tell your language guarantees.}

\newpage
\ifusenix
    \bibliographystyle{plain}
\fi
\bibliography{bibliography}


\appendix

\section{Number of packages converting values from C/C++ types}\label{app:numpack}

\input{figs/bar_types}

In Figure~\ref{fig:types_package}, we depict the total number of packages that explicitly convert values to a given type.
Casting to object, number or string types are the most prevalent conversions, performed by 77\% of the packages. 
The relatively low number of functions (16.3\%) shows that at most one in three packages perform non-blocking, asynchronous operations ($M_{12}$), as these operations require a function object as argument to be invoked upon completion.

\section{Example cross-language data-flow graphs}\label{sec:app}
In this section, we provide examples for cross-language data-flow graphs for different scenarios.
Additionally, we provide more complex examples in the supplementary material of this paper\footnote{\url{https://native-extension-risks.herokuapp.com}}.

In Figure~\ref{fig:zopfli}, we provide a small example unsanitized flow from the \texttt{buffer} parameter in JavaScript to the type conversion API \texttt{Buffer::Data(*)} in C++ in package \texttt{zopfli-node}, version 2.0.3.
In this figure, it can be observed that the \texttt{buffer} parameter flows into  \texttt{Buffer::Data(*)} in C++ with no sanitization.

In Figure~\ref{fig:iltorb}, we provide an example sanitized flow from the \texttt{input} parameter to the type conversion API  \texttt{Buffer::Data(*)} in C++ in package \texttt{iltorb}, version 1.0.0.
We mark the node where sanitization takes place with red and note that in this example, the sanitization is carried out in the JavaScript front-end.

In Figure~\ref{fig:bcrypt}, we provide a somewhat complex sanitized flow from the \texttt{rounds} parameter to the type conversion API  \texttt{Int32Value()} in C++ in package \texttt{nan-bcrypt}, version 0.7.7.
We mark the nodes where sanitization takes place with red and note that in this example, sanitization is carried out both in the JavaScript front-end and in the C++ native extension code.

\input{figs/appedix}

\section{False positive produced by \toolname{}}\label{sec:app2}
In this section, we provide additional details about the false positive detected by \toolname{}. In Listing~\ref{lst:fp} we show the relevant source code for the flagged application. The application processes seven attributes of the post request's body by saving them in the \texttt{sqlite3} database. To deploy our payload identified in Section~\ref{subsec:exploits}, we need to pass a specially-crafted \textbf{object} as an entry in the array passed as the second argument to \texttt{sqlite3.run()}. The type conversions in line 5, 6, 7, 8, and the type checks in line 27, 34, 41 ensure that only numbers and strings are allowed, respectively. The presented false positive is caused by limitations in our current prototype, and not by fundamental shortcomings of the presented approach.

\begin{figure*}[p!]
\begin{lstlisting}[caption={False positive produced by \toolname{}. The type conversions in line 5, 6, 7, 8, and the type checks in line 27, 34, 41 prevent the deployment of remote type confusion payloads.}, label=lst:fp]
module.exports = (db) => {
    app.get('/health', (req, res) => res.send('Healthy'));

    app.post('/rides', jsonParser, (req, res) => {
        const startLatitude = Number(req.body.start_lat);
        const startLongitude = Number(req.body.start_long);
        const endLatitude = Number(req.body.end_lat);
        const endLongitude = Number(req.body.end_long);
        const riderName = req.body.rider_name;
        const driverName = req.body.driver_name;
        const driverVehicle = req.body.driver_vehicle;

        if (startLatitude < -90 || startLatitude > 90 || startLongitude < -180 || startLongitude > 180) {
            return res.send({
                error_code: 'VALIDATION_ERROR',
                message: 'Start latitude and longitude must be between -90 - 90 and -180 to 180 degrees respectively'
            });
        }

        if (endLatitude < -90 || endLatitude > 90 || endLongitude < -180 || endLongitude > 180) {
            return res.send({
                error_code: 'VALIDATION_ERROR',
                message: 'End latitude and longitude must be between -90 - 90 and -180 to 180 degrees respectively'
            });
        }

        if (typeof riderName !== 'string' || riderName.length < 1) {
            return res.send({
                error_code: 'VALIDATION_ERROR',
                message: 'Rider name must be a non empty string'
            });
        }

        if (typeof driverName !== 'string' || driverName.length < 1) {
            return res.send({
                error_code: 'VALIDATION_ERROR',
                message: 'Rider name must be a non empty string'
            });
        }

        if (typeof driverVehicle !== 'string' || driverVehicle.length < 1) {
            return res.send({
                error_code: 'VALIDATION_ERROR',
                message: 'Rider name must be a non empty string'
            });
        }

        var values = [req.body.start_lat, req.body.start_long, req.body.end_lat, req.body.end_long, req.body.rider_name, req.body.driver_name, req.body.driver_vehicle];
        
        const result = db.run('INSERT INTO Rides(startLat, startLong, endLat, endLong, riderName, driverName, driverVehicle) VALUES (?, ?, ?, ?, ?, ?, ?)', values, function (err) {
            if (err) {
                return res.send({
                    error_code: 'SERVER_ERROR',
                    message: 'Unknown error'
                });
            }
        ...
        });
    ...
    };
}
\end{lstlisting}
\end{figure*}

\section{Sinks and sanitizers used by our prototype}\label{sec:app3}
{Below, we enumerate the sinks and sanitizers used by our prototype for detecting missing type checks. For conciseness, we use the metavariable {\color{blue}\#type\#} for specifying several sinks or sanitizers from the same family, e.g., instead of mentioning both  \texttt{napi{\textunderscore}get{\textunderscore}value{\textunderscore}int32()} and \texttt{napi{\textunderscore}get{\textunderscore}value{\textunderscore}string{\textunderscore}utf8()}, we use the notation  \texttt{napi{\textunderscore}get{\textunderscore}value{\textunderscore}{\color{blue}\#type\#}()}  to refer to all APIs of that form.}

Set of sinks (all in C/C++):
\begin{itemize}
    \item \texttt{napi{\textunderscore}get{\textunderscore}buffer{\textunderscore}info()} 
    \item \texttt{Buffer::Data()} 
    \item \texttt{Buffer::Length()} 
    \item
    \texttt{*.As{\textless}{\color{blue}\#type\#}{\textgreater}} 
    \item
    \texttt{*.To{\textless}{\color{blue}\#type}{\#\textgreater}}
    \item
    \texttt{*.To{\color{blue}\#type\#}()} 
    \item \texttt{*.ToLocalChecked()}
    \item \texttt{*::Cast()}
    \item 
    \texttt{napi{\textunderscore}get{\textunderscore}value{\textunderscore}{\color{blue}\#type\#}()} 
    
\end{itemize}

Set of sanitizers:
\begin{itemize}
    \item (C/C++) \texttt{napi{\textunderscore}is{\textunderscore}{\color{blue}\#type\#}()} 
    \item (C/C++) \texttt{napi{\textunderscore}typeof()} 
    \item (C/C++) \texttt{Nan::Check()} 
    \item (C/C++) \texttt{*.HasInstance()}
    \item (C/C++)
    \texttt{*.Is{\color{blue}\#type\#}()} 
    \item (JavaScript) \texttt{typeof} 
    \item (JavaScript) \texttt{Buffer.isBuffer()}
\end{itemize}

\section{Complete list of misuses we identified}
\label{ap:list-all}
In Table~\ref{fig:crashes} we show the list of npm packages for which we could trigger a hard crash, their reach in the ecosystem, and the API we used for the exploit. We remind the reader that we use the hard crashes as a testing oracle, indicating a potential security problem.
Some of these packages do not compile with the latest Node.js version and require a legacy version of the runtime instead. Others require a specific library on the operating system before installation. On our setup we could, however, meet such strict constraints by acting on the compilation error we observed on unsuccessful installation attempts. We reported all these issues to the package maintainers.

\begin{table}
\setlength{\tabcolsep}{3pt}
    \caption{Npm packages in which we identified a previously unknown hard crash. The endpoint represents the package's method that we use for the proof of concept. With \#main\# we depict the default method exposed by the package.}
    \label{fig:crashes}
  \rowcolors{2}{gray!25}{white}
    \centering
    \resizebox{0.9\columnwidth}{!}{%
    \begin{tabular}{c|c|c|c}
         \rowcolor{gray!40}
         \textbf{Package name} & \textbf{Version} & \textbf{Reach\footnote{As defined by Zimmermann et al.~\cite{ZimmermannSTP19}.}} & \textbf{Endpoint}\\
         \hline
@alien.sh/signals & 1.0.0 & 1 & Register\\
bigint-hash & 0.2.2 & 3 & update\\
bigint-buffer & 1.1.5 & 97 & toBigIntLE\\
bignum	& 0.13.1 & 168 & powm\\
binary-diff & 1.0.0 & 1 & \#main\#\\
bkjs-utils & 0.2.8 & 1 & getUser\\
ced & 0.0.1 & 3 & \#main\#\\
csac-ed25519 & 0.0.3 & 1 & Verify\\
csocket	& 1.0.3 & 1  & send\\
cuckaroo29b-hashing & 1.0.0 & 1 & cuckaroo29b\\
fast-string-search & 1.4.1 & 1  & indexOf\\
gs-node-lmdb & 0.9.0 & 1 & Cursor\\
int64-napi & 1.0.1 & 3 & divide\\
jitterbuffer & 0.1.14 & 3 & put\\
libasar{\textunderscore}enc & 1.0.0 & 1 & \#main\#\\
libxmljs & 0.19.7 & 237 & parseXml\\
multi-hashing & 1.0.0 & 10 & scryptjane\\
node-crc & 1.3.0 & 4  & crc64iso\\
node-lzma & 0.1.0 & 1  & compress\\
pg-native & 3.0.0 & 92 & query\\
libpq & 1.8.9 & 4 & {\$}execParams\\
node-mbus & 1.2.1 & 3 & \#main\#\\
phin-ecdh & 1.0.0 & 1  & encrypt\\
pixel-change & 1.0.0 & 2  & \#main\#\\
rapid-crc & 1.0.10 & 2 & crc32c\\
roaring	& 1.0.6 & 6 & \_initTypes\\
sbffi & 1.0.4 & 1 & getNativeFunction\\
sendto & 1.0.3 & 1 & \#main\#\\
sqlite3	& 5.0.1 & 1905 & run\\
termios-fixedv12 & 0.1.9 & 2 & getattr\\
time & 0.12.0 & 56 & setTimezone\\
utf-8-validate & 5.0.8 & 551 & \#main\# \\
zopfli-node & 2.0.3 & 1 & deflateSync
    \end{tabular}
}
\end{table}

\section{Performance of graph extraction}
\label{ap:perf}
We run all our experiments on a server with 64 AMD EPYC 7H12 cores, 2TB of memory, and running Debian 11. In Figure~\ref{fig:perf_analysis}, we show the time used by our prototype to extract the JavaScript and the C/C++ graphs. For 75\% and 93\% of the packages, the analysis finishes in less than 10 seconds for native and JavaScript parts, respectively. This shows that that the proposed approach scales well and that packages with native extensions tend to have more complex C/C++ code than JavaScript. For \pkgsTimeoutC{} packages, the graph extraction times out while analyzing the C/C++ part, while for \pkgsTimeoutJS{} packages it does so for the JavaScript part. Since this represents less than 0.5\% of the considered packages, we deem it is a minor shortcoming of our current prototype, and not a fundamental limitation of the proposed methodology. 

\begin{figure}[t]
\input{perf}
\end{figure}

\section{Manual analysis procedure}
\label{ap:manual}
During our manual analysis, we first verify if there is a JavaScript check that protects the reported vulnerable endpoint (\say{type check in JavaScript} in Figure~\ref{fig:typecheck}). We then verify if indeed there is an unsanitized flow in the C/C++ part, i.e., that there are no method calls that perform sanitization that were missed by our intra-procedural analysis (\say{false positives}). We then proceed with the installation of the given package on our machine, which turned out to be a very challenging task. We were unable to install the majority of the reported packages (\say{unable to verify}) due to several reasons: legacy code not running in our considered Node.js runtimes, missing hardware, different operating system, missing installed libraries. To maximize the number of packages that we can install, we attempt installing with five different Node.js versions: 15.4.0, 14.15.0, 12.22.1, 8.17.0, 0.12.18. For the packages that we could install, we attempt to write an exploit that produces a hard crash (\say{exploited}). If we fail to do so, we reanalyze the code and assign it to one of the other categories mentioned earlier.

\end{document}

%% file: introduction.tex
\section{Introduction}
Originally, the main use case for modern scripting languages~\cite{ousterhout1998scripting} like Python, Ruby or JavaScript was the development of web applications.
Recently, though, they became tremendously popular, general-purpose programming languages with powerful emerging use cases like TensorFlow for machine learning in Python or Electron.js for portable desktop applications in JavaScript. 
This development is supported by massive open-source ecosystems 
such as npm, PyPI and RubyGems. There is a large body of work identifying a plethora of security risks that affect these software repositories~\cite{ZimmermannSTP19, DecanMC17, DavisCSL18, StaicuPL18,duantowards,AbdalkareemNWMS17}, which in turn, can impact real-world websites~\cite{StaicuP18}. However, prior work only considers security risks present in the scripting code, thus, ignoring the important cross-language interactions in these ecosystems.

Native extensions are a convenient way to allow low-level functionality, 
to be directly invoked from a scripting language. Package managers like \texttt{npm}, \texttt{pip} or \texttt{gem} enable smooth usage of such extensions by compiling at install time the extension's binary~\cite{NpmNE,PipNE,GemNE}. At runtime, the binary is loaded on-demand in the scripting code's process, unlocking cross-language cooperation. 
In this way, developers can  expose hardware capabilities that were originally beyond the reach of the scripting language. 
Native extensions also enable the reuse of mature, legacy code written in  low-level languages. Databases like SQLite\footnote{\url{https://www.sqlite.org}} or cryptographic libraries like OpenSSL\footnote{\url{https://www.openssl.org/}} are often exposed using native extensions. 
Finally, native extensions aid the development of performance-critical code, in low-level languages. For example, a non-negligible part of TensorFlow is written in C++ and exposed to Python through bindings.

\input{figs/examplenpm}

One can think of native extensions as the democratization of binding layer, which glues the language engines with their surrounding environment. 
Previous work~\cite{BrownNWEJS17,dinh2021favocado} discusses the security risks incurred by this layer and provides evidence that vulnerabilities are prevalent even in binding code of popular runtimes like Node.js. This type of code is usually developed by highly-skilled developers, whereas native extensions can be written by anyone. 
As one may expect, writing reliable native extensions is difficult since subtle bugs may arise at the language boundary. The main culprit for this are the fundamental differences in representing data in the two languages, e.g., weak dynamic typing vs. strong static typing. 
Moreover, a mistake in an extension may propagate in the ecosystem, affecting several libraries that depend on it, or even compromising production-ready applications.

Let us consider the example in Figure~\ref{fig:example} to illustrate how bugs may arise when using native extensions. 
The \texttt{nativepad} package uses a native extension to pad a given string to the right with the literal \texttt{"pad"}. Its native extension in Figure~\ref{fig:extension-example} employs four calls to the extension API: one at line~5 to retrieve the arguments, one at line~7 to get the length of the first argument, one at line~10 for converting the JavaScript string into a C one, and finally one call at line~12 to convert the C string back into a JavaScript one. Additionally, the extension allocates the memory to store the padded string, and performs the string concatenation using \texttt{strcat}. The JavaScript code of the \texttt{nativepad} package in Figure~\ref{fig:wrapper-example} is trivial, performing a simple null check on the input and invoking the \texttt{Pad} function of the native extension. Now, let us consider a client in Figure~\ref{fig:client-example} that invokes the exported function with different arguments. Note that the client is oblivious to the use of native extensions, i.e., the \texttt{require} statement in line~1 would be exactly the same for loading a ``pure'' JavaScript package. 
When invoking \texttt{nativepad} with a well-behaved string, the padding is performed as expected. However, when the null terminator (\texttt{\textbackslash0}) is present in the string, the native extension exposes uninitialized memory. By the JavaScript runtime this character is treated as any other character, i.e., counting it towards the string length, while it leads to string termination in C. 
Even more surprising behavior emerges, e.g., hard crash of the Node.js process, if unexpected values (e.g., Booleans or certain object literals) are provided.
Such outcomes may surprise users, potentially leading to security incidents, e.g., denial of service.


While the considered example does not follow the best practices of the native extension API, e.g., checking the argument type or the return value
, we believe that the runtime should be robust enough to protect against such a \emph{misuse}. We notice that there is a large design space for a native extension API and that different design decisions make programming with the obtained API more dangerous than others. To explore this design space, we study the native extension API in three popular scripting languages and show that misuse is possible in each of them, though, there are important differences across languages. The Node.js API is by far the most permissive, allowing several types of misuse, such as calling a native extension with insufficient arguments or integer overflow for numeric values exchanged across the language boundary. 

To study the security implications of using native extensions, our methodology first identifies misuse in open-source libraries. To that end, we perform both intra-procedural and cross-language static analysis. We propose a simple, yet effective way of constructing cross-language graphs that combines the two functions that are  closest to the language boundary. We then perform demand-driven data-flow analysis on open-source web applications to study the impact of the library-level problems at application level.

In our evaluation, we first perform an
empirical study of the prevalence of misuse in \nbPkgsStudy{} npm packages with native extensions. We show that even popular packages are prone to misuse and we provide evidence that an attacker can cause real harm to web applications by leveraging the bugs introduced by API misuse. Concretely, we provide proof-of-concept exploits for \numCrashes{} npm packages, showing that by manipulating the inputs to these packages (strong attacker model), attackers can break the language guarantees. Moreover, we identify \nbVulnWapps{} open-source web applications in which hard crashes can be caused remotely (weak attacker model).  We were assigned \nbCVEs{} CVEs for our findings, most labeled as high severity.

In summary, we provide the following novel contributions: 
\begin{itemize}[noitemsep,topsep=0pt]
    \item We are the first to analyze in detail the security risks of native extensions in scripting languages. Several design decisions enable  
    vulnerabilities and burden the developer with the task of using the API in a secure way.
    \item We present a novel methodology that enables the study of vulnerabilities caused by misuse of the native extension API. We show how cross-language static analysis can be used for automatic vulnerability detection.
    \item We provide evidence that vulnerabilities caused by native extensions are present in open-source software packages and that they also affect web applications using them.
\end{itemize}

%% file: figs/examplenpm.tex
\begin{figure*}
\center
\centering
    \begin{subfigure}{0.43\textwidth}
    \center
    \centering
    \caption{JavaScript client of the package \texttt{nativepad}}
    \label{fig:client-example}
    \captionsetup{position=top}
    \begin{lstlisting}[language=JavaScript,numbers=left]
let nlib = require('nativepad');
nlib('foo/*#\mytikzmark{c1}#*/'); //returns "foopad"
nlib('foo\0bar/*#\mytikzmark{c2}#*/'); //returns "foo" followed by three uninitialized bytes
nlib(true/*#\mytikzmark{c3}#*/); //returns 4 uninitialized bytes
nlib({toString: /*#\mytikzmark{c4}#*/42}); //hard crash (segfault)
    \end{lstlisting}
    
    \caption{JavaScript code for the package \label{fig:wrapper-example}
\texttt{nativepad}}
    \captionsetup{position=top}
    \begin{lstlisting}[language=JavaScript,numbers=left]
let addon=require('bindings')('addon.node');
module.exports = (str/*#\mytikzmark{n0}#*/) => {
    if (!str)
        throw 'Invalid string';
    return addon.Pad(str/*#\mytikzmark{n1}#*/);
}
    \end{lstlisting}

    \end{subfigure}
    \begin{subfigure}{0.55\textwidth}
    \caption{C++ code for the native extension}
    \captionsetup{position=top}
    \label{fig:extension-example}
    \begin{lstlisting}[language=C++,numbers=right,emph={int    ,char,double,float,unsigned, napi_value, napi_env, napi_callback_info,size_t},keywords={assert,sizeof,return},emphstyle={\color{solablue}}]
napi_value Pad(napi_env env, napi_callback_info info) {
   napi_status status;
   size_t argc = 1, strSize;   
   napi_value args[1], result;
   status = napi_get_cb_info(env, info, &argc, args, NULL, NULL);   
   assert(status == napi_ok);
   napi_get_value_string_utf8(env, /*#\mytikzmark{n2}#*/args[0], NULL, NULL, &strSize);   
   strSize = strSize + 4;
   char myStr[strSize];
   napi_get_value_string_utf8(env, /*#\mytikzmark{n3}#*/args[0], myStr, strSize, NULL);   
   strcat(myStr, "pad");
   napi_create_string_utf8(env,myStr,strSize,&result);
   return /*#\mytikzmark{n4}#*/result;
}
    \end{lstlisting}
    \begin{tikzpicture}[->,>=stealth', remember picture,overlay, dashed]
   \path[line width=0.6pt,opacity=0.2] ([yshift=2mm, xshift=-1.5mm]n1) edge [bend left=15] ([yshift=2.5mm, xshift=2.5mm]n2);
    \path[line width=0.6pt, opacity=0.2] ([yshift=2mm, xshift=-1.5mm]n1) edge [bend left=15] ([yshift=2.5mm, xshift=2.5mm]n3);
    \path[line width=0.6pt, opacity=0.2] ([yshift=-1mm, xshift=2.5mm]n4) edge [bend left=5] ([yshift=-1mm, xshift=-10mm]n1);
    \path[line width=0.6pt, opacity=0.2] ([yshift=-1mm, xshift=2.5mm]c1) edge [bend left=15] ([yshift=2mm, xshift=-2mm]n0);
    \path[line width=0.6pt, opacity=0.2] ([yshift=-1mm, xshift=2.5mm]c2) edge [bend left=15] ([yshift=2mm, xshift=-2mm]n0);
    \path[line width=0.6pt, opacity=0.2] ([yshift=-1mm, xshift=2.5mm]c3) edge [bend left=15] ([yshift=2mm, xshift=-2mm]n0);
    \path[line width=0.6pt, opacity=0.2] ([yshift=-1mm, xshift=2.5mm]c4) edge [bend left=15] ([yshift=2mm, xshift=-2mm]n0);
   \end{tikzpicture}
    \end{subfigure}
    
    \caption{Example of a hypothetical npm package  called \texttt{nativepad} (b), its native extension (c) and a client invoking it (a). The dashed arrows show the data flows between the three components.}
    \label{fig:example}
\end{figure*}
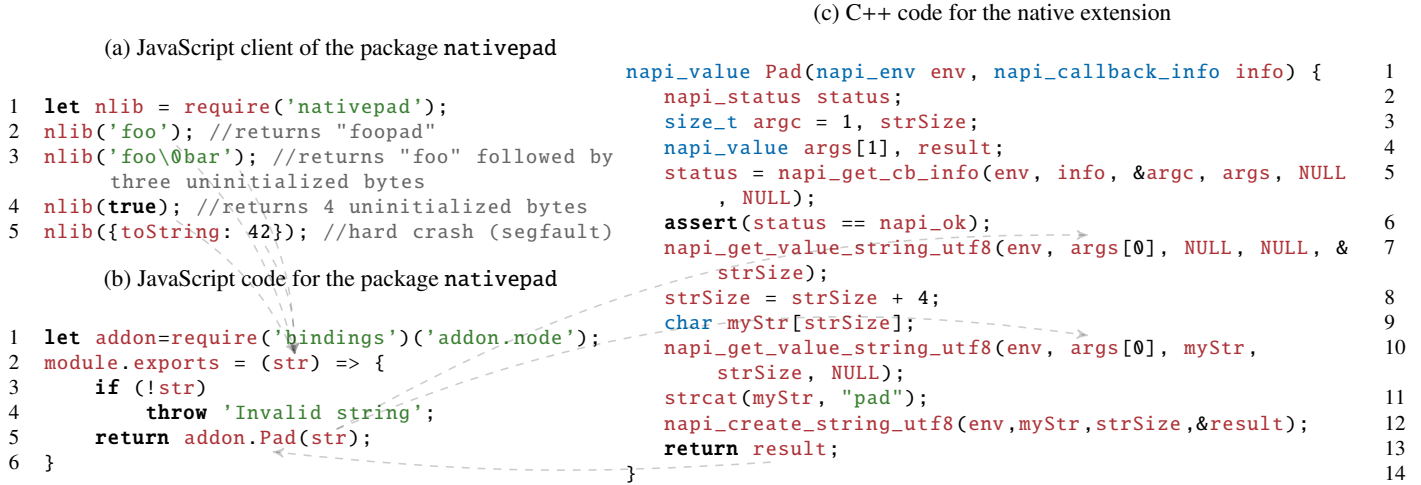

%% file: threat_model.tex
\section{Threat Model}
\label{sec:threatmodel}

We assume that attackers have neither control over the native extension's code, nor privilege to execute arbitrary scripting language code. Developers of extensions are not malicious, but they might inadvertently introduce vulnerabilities in their code. We consider a native extension to be vulnerable if it can be used in a way that breaks the guarantees of the scripting language. For example, if it can crash the process, read/write to unintended locations, or execute arbitrary code. Note that native extensions may also be used to hide malicious payloads in supply chain~\cite{duantowards} or protestware~\cite{protestware} attacks. Our goal is to demonstrate the dangers of native extension vulnerabilities even if the developers are honest. Therefore, detecting supply chain abuse attacks are out of scope for this work.

We propose two attacker models. First, for analyzing packages with native extensions in isolation, we assume a \textbf{strong attacker model}, in which attackers are able to control any argument passed to a library with native extensions, as well as their number. However, only objects that can be serialized as JSON are allowed as arguments and no modifications of the builtins are permitted. For example, attacker-defined functions or modifications of \texttt{Object.prototype} are not allowed in our setting. Thus, even this strong attacker model is much weaker than the one used in prior work on JavaScript bindings\cite{BrownNWEJS17,dinh2021favocado}, where the authors assumed that attackers can inject arbitrary code in the engine. 
Ultimately, we aim to identify vulnerabilities within native extensions that can be triggered remotely, under a \textbf{weak attacker model},
where we assume a web attacker that can only provide inputs to a web application through its HTTP interface. These inputs may propagate to a native extension and trigger a vulnerability in its implementation. 

Package managers for scripting languages exhibit a significant amplification effect for vulnerabilities~\cite{MastrangeloPMLH15, EvansCS20,ZimmermannSTP19}. We note that this effect also increases the significance of native extension vulnerabilities. Since package managers do not have any privilege restriction in place or a transparency mechanism to warn about the usage of security-sensitive APIs, they may transitively depend on libraries with native extensions, without being aware of this fact. Attackers, thus, can exploit vulnerabilities caused by native extensions in client packages. 



%% file: figs/overview.tex
\tikzset{
  basic/.style  = {draw, text width=2.5cm, minimum height=1cm, rectangle, text centered, fill=solablue!10},
  noframe/.style  = {text width=2.5cm, minimum height=1cm, rectangle, text centered},
}

\begin{figure*}[!ht]
\centering
\includegraphics[width=0.92\textwidth]{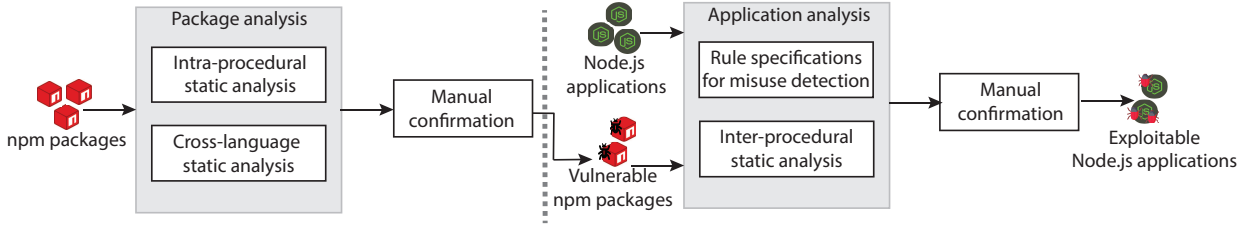}
 	\caption{Overview of our methodology for identifying native extension vulnerabilities and for studying their impact.}
 	
 	\label{fig:framework_v2}
\end{figure*}

%% file: methodology.tex
\section{Methodology}
\label{sec:methodology}

While native extensions can be directly integrated in (web) applications, we believe that it is more common for these extensions to be first encapsulated in a package. 
Hence, we propose two levels of static analysis to detect native API misuse vulnerabilities. We depict our analysis pipeline for Node.js and npm in Figure~\ref{fig:framework_v2}, but we believe that it can be easily adapted for other scripting languages and their ecosystems. First, we run a package-level analysis to detect vulnerable npm packages due to insecure native extensions, under the strong attacker model. We propose running both simple, intra-procedural analyses, but also cross-language ones. Specifically, we create a common representation for both C/C++ and JavaScript code present at the language boundary to detect problematic native extensions within a package. After finding a vulnerable package, we use inter-procedural backward data-flow analysis to study its impact on applications that use the package, under the weak attacker model assumption.

\subsection{Package analysis}
\label{sec:pkg-analysis}

Since most of the  native extensions we encountered are relatively small, and many misuses can be formulated as flow problems, we propose specifying the misuse detection as a graph traversal problem on the data-flow graph. 
However, as we show in Section~\ref{sec:evaluation}, this may lead to a significant number of false positives because the analysis does not have information about how data is handled in the upper layer, i.e., in the scripting language. Hence we also propose unifying the data-flow graphs of the two languages.

\textbf{Intra-procedural analysis.}
The first step of our analysis is to create a data-flow graph of the target functions. Our definition for such a graph is very permissive: nodes $N$ represent lines of the program and edges $E$ depict explicit information flows. For instance, the green part of Figure~\ref{fig:example_dot} shows the data-flow graph for the example in Figure~\ref{fig:extension-example}. The nodes represent statements and the edges represent data flows between them. A slightly different representation is the green part of Figure~\ref{fig:bcrypt} in Appendix~\ref{sec:app}, where some of the nodes are only partial statements. We argue that the exact representations may vary as long as the semantics of the edges are preserved.

We then associate special meaning to particular nodes in the graph. $n_0$ is the root node of the graph where the traversal starts from, corresponding to the method definition statement in the source code. Thus, it has outer edges towards all the nodes in which parameters are referenced. $S$ is the list of sink nodes that the analysis is interested in, e.g., the \texttt{Buffer::Data()} call in Figure~\ref{fig:bcrypt} for a wrong argument type vulnerability ($M_3$). $\overline{S}$ is the list of sanitizers that invalidate a given flow to the sink. 

\input{figs/example_dot}

Our analysis reports a security vulnerability iff:

\begin{itemize}[noitemsep,topsep=-\parskip]
    \item $\exists \, s \in S $ such that $ n_0 \rightsquigarrow s$,
    \item $\nexists \, \overline{s} \in \overline{S} $ such that $n_0 \rightsquigarrow \overline{s}$,
\end{itemize}

where $a \rightsquigarrow b$ represent a path from $a$ to $b$ on the graph. We note that the presented analysis is not argument-sensitive, if any data flow to the sanitizer is detected, the flow to the sink is considered safe. This  is a pragmatic design decision that can lead to many false negatives in practice. Nevertheless, in this work we do not aim for a complete solution to the described problem, but for showing the feasibility of an automated detection technique in this domain.

\textbf{Cross-language analysis.}
We observe that many relevant API calls, e.g., sanitizers, happen in the two functions that are closest to the language boundary: one in JavaScript and one in C/C++. For identifying such pairs, we search for calls to the native extension API that map low-level functions to their high-level names. All the considered APIs in Section~\ref{sec:study} require such calls during the initialization of a native extension. Let us assume we want to expose the \texttt{Foo} function from C/C++ to the scripting language, with the name \texttt{"foo"}. The syntax used by the considered APIs for binding the two entities is:
\begin{lstlisting}
// Node.js-Nan
Set(module, New<v8::String>("foo"),
      New<v8::FunctionTemplate>(Foo));
// Node.js-N-API
napi_define_properties(...,{"foo",...,Foo,...}});
// Ruby
rb_define_method(module,"foo",Foo,1);
// Python
PyModule_Create({...,{"foo",(PyCFunction)Foo},...});
\end{lstlisting}
Once we identify this mapping, we merge the data-flow graphs of the two functions by adding an edge from the node in the JavaScript graph corresponding to the native extension call, to the definition node of the invoked C/C++ function. 
Finally, we perform the same analysis described above, on the obtained cross-language graph.

Let us consider  Figure~\ref{fig:example_dot} that shows the cross-language data-flow graph corresponding to the native extension in Figure~\ref{fig:example}. We assume that we are interested in detecting unchecked type conversions. By applying our analysis starting from $JS2$, we can detect a path to $C7$ that corresponds to a call to \texttt{napi\_get\_value\_string\_utf8()}, i.e., the sink. Since there is no statement either in JavaScript nor in C/C++ that checks that the type of the argument is string (sanitizer), the analysis produces a warning for this case. Let us assume that in line 3 of Figure~\ref{fig:wrapper-example} there is a type check instead of a null check.
{Our cross-language analysis is path-insensitive, i.e., if we detect a flow from the source to a sanitizer, we consider the usage safe. 
Therefore,
}
the analysis would detect a flow to the sanitizer marked with a dashed red circle, i.e., in $JS3$, and would not produce a warning.

\textbf{Implementation details.}
For extracting the data-flow graphs, we use Joern~\cite{YamaguchiGAR14} for C/C++ files and Google Closure Compiler~\cite{closure} for JavaScript. We instruct Joern to output the code property graph as a dot\footnote{
\url{https://en.wikipedia.org/wiki/DOT_(graph_description_language)}} file and further pre-process it, by only preserving the data-flow edges. We also add edges from the function definition node, i.e., the first node, to the nodes accessing the \texttt{info[*]} and \texttt{args[*]} objects, which are the arguments coming from JavaScript. Joern fails to detect these edges, because the arguments do not appear verbatim in the function declaration. For the Google Closure Compiler, on the contrary, we build our custom compiler pass to extract def-use pairs from its internal representation and output them in a dot file. We run both Joern and the Closure-based analysis with a budget of 15 minutes per analyzed package.

For finding the two functions at the language boundary, we first perform a simple static analysis of the JavaScript code to detect which of the exposed C/C++ functions are called directly and in which JavaScript function. We then proceed by resolving these calls by analyzing the C/C++ code and identifying API calls to functions such as \texttt{napi\_define\_properties} described above. 
Once we identified the two functions, we retrieve their corresponding dot representations and merge them as described earlier. We then analyze the obtained graph and output security violations. 
Thereafter, we manually verify each security violation by attempting to exploit the misuse through the package's API. In case of success, we proceed to study the vulnerability's impact on real-world web applications. Additionally, whenever we identify an exploitable violation, we proceed to manually look for other misuses in that package, under the assumption that misuses tend to occur together.

\textbf{Security Modelling.} Our current prototype is targeted towards studying an important subset of the misuses identified in Table~\ref{tab:comparative-analysis}: missing type checks ($M_{3}$, $M_{4}$). For $M_{3}$, we specify the list of sinks based on the APIs we studied in Section~\ref{sec:study}, and the list of sanitizers based on idiomatic type checks in the two languages, together with the APIs provided by N-API and Nan for type checking. We provide the complete list of sinks and sanitizers in Appendix~\ref{sec:app3}. For $M_4$, we additionally consider checks on number of arguments as sanitizers. The supplementary manual analysis step described above allows us to identify misuses that go beyond ($M_{3}$, $M_{4}$), once an initial missing type check was identified in a given package.


\subsection{Application analysis}\label{sec:app-analysis} 
{The existence of native extension vulnerabilities in npm packages motivated us to investigate their impact on Node.js web applications. Specifically, after manual confirmation of the vulnerabilities found in Section~\ref{sec:pkg-analysis}, we study their exploitability in the web application context. We formulate the detection of web applications using vulnerable packages as a flow problem, which can be automatically detected. To make our analysis scalable, we chose to use static analysis over dynamic analysis. This is because dynamic analysis requires running an application to monitor its runtime behavior~\cite{DBLP:conf/sigsoft/SenKBG13, DBLP:journals/csur/AndreasenGMPSSS17, dinh2021favocado}. Manually setting up and running a diverse set of Node.js web applications with their heterogeneous software and library dependencies are infeasible. 


Therefore, we build a new demand-driven, def-use based, static data-flow analysis framework for JavaScript, named \toolname{}, for our needs.
Finally, we use \toolname{} on Node.js applications to detect exploitable uses of insecure native extensions.} It is worth noting that demand-driven data-flow analysis has already been proven to effectively detect various kinds of API misuse in other languages~\cite{RahamanXASTFKY19, BianchiFMKVCL18, SpathDAB16, GensSDS18}. 
 In this section, we discuss different components of our \toolname{} framework. Note that \toolname{} guarantees neither soundness (i.e., absence of bugs), nor completeness (i.e., absence of false alarms). However, like other practical static analysis tools, it favors completeness and efficiency over soundness. {This design choice is acceptable for our use case, since our goal is to run~\toolname{} scalably on a large number of web applications.}
 
\textbf{Rule specification.} 
 {Our analysis takes rule specifications as input, which are manually created for a given vulnerable native extension API.}
 A rule specification contains the API of interest and a callback function to check its misuses. The API definition consists of the function name and the parameter of interest. For example, to detect unsanitized inputs from the network to the function \texttt{run(query, data)} of the \texttt{sqlite3} package, one might specify the rule as follows.
\setlength{\abovedisplayskip}{3pt}
\setlength{\belowdisplayskip}{3pt}
 \begin{equation*}
 IsMisuse \frac{{p_{o} : \textit{run(\_, \textbf{data})}, ~~ P : \{p_{i}\}, \text{ s.t. } p_{i} \rightsquigarrow p_{o}, \forall i \in [1, |P|]}}{\text{if \texttt{req} \textbf{or} \texttt{req.body} }\epsilon ~P \text{ then \textbf{true} else  \textbf{false}}}
\end{equation*}
 
 Here, $IsMisuse$ is the callback function that takes $P$ as input and outputs \textit{true} if a misuse is found and \textit{false} otherwise. $p_{o}$ is the API definition, and $P$ is the set of all unsanitized influences on $p_{o}$. \texttt{req} is the object containing the request data to the server. $\rightsquigarrow$ represents the \textit{direct influence} on $p_{o}$. Our demand-driven analysis starts from the API invocation to find all program entities that influence it. 
 
 \noindent
 \textbf{Intra-procedural backward data-flow analysis.} An analysis to find the data flows to a given program point (invocations of the defined APIs) is known as \textit{backward data-flow analysis}. We build our intra-procedural backward data-flow analysis on top of Google Closure Compiler's internal data-flow analysis framework. Closure's data-flow analysis framework provides an implementation of the worklist algorithm~\cite{DBLP:conf/pldi/PadhyeK13, RahamanXASTFKY19}. To calculate data flows at a given program point, we implemented a flow-insensitive def-use analysis by using the Abstract Syntax Tree~(AST) representation of the code provided by the Closure Compiler. \textit{Definition} (in short, \textit{def}) of a variable \textit{x} is an instruction that writes to \textit{x}. \textit{Use} of a variable \textit{y} is an instruction that reads \textit{y}. An analysis which utilizes the def-use relationship of variables is known as \textit{def-use analysis}. Our def-use analysis collects direct influences and avoids any orthogonal function invocations. This is because we use \toolname{} to find \textit{raw inputs} from the network or file system (Section~\ref{subsec:webapps}), where processing is typically performed with orthogonal function calls.
 
 \noindent
 \textbf{Call-graph generation.} 
 We implement our call-graph generator on top of Google Closure's AST traversal algorithm. We traverse the AST to find function \textit{definition} and \textit{invocation} nodes and collect all the caller-callee relationships within a JS file. We represent anonymous function definitions with their line number and the starting position. For this prototype implementation we do not handle function aliasing.
 
 \noindent
 \textbf{Inter-procedural backward data-flow analysis.}
Our analysis starts by finding all the call sites of the provided API invocations. It then runs intra-procedural backward data-flow analysis on all the caller functions by using the given call sites. We run the process recursively by following the caller-callee chain upward. Then, we stitch all the intra-procedural data-flow summaries to form inter-procedural data-flow results. Finally, \toolname{} invokes the rule-specific callback functions on these 
results to find misuses. 

%% file: figs/example_dot.tex
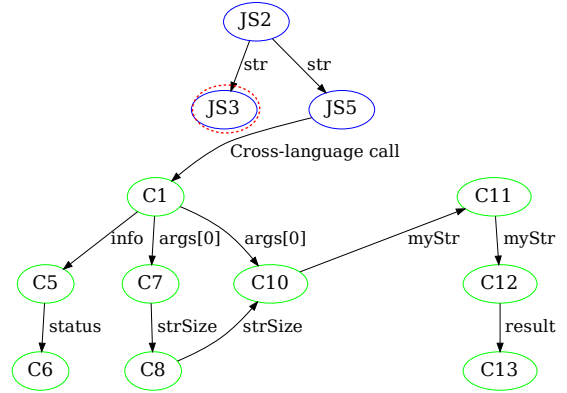
\begin{figure}[t]
    \centering
    \resizebox{\columnwidth}{!}{
\begin{tikzpicture}
\node[anchor=south west,inner sep=0] at (0,0) {\digraph{ex}{
ratio=0.7;
node [fontsize=22];
edge [ fontsize=20 ];
  "JS2" [label ="JS2",color=blue]
  "JS3" [label ="JS3",color=blue]
  "JS5" [label ="JS5",color=blue]
  "C1" [label ="C1",color=green]
  "C5" [label ="C5",color=green]
  "C6" [label ="C6",color=green]
  "C7" [label ="C7",color=green]
  "C8" [label ="C8",color=green]
  "C10" [label ="C10",color=green]
  "C11" [label ="C11",color=green]
  "C12" [label ="C12",color=green]
  "C13" [label ="C13",color=green]
  "JS2" -> "JS3" [label="str"]
  "JS2" -> "JS5" [label="str"]
  "JS5" -> "C1" [label=" Cross-language call"]
  "C1" -> "C10" [label="args[0]"]
  "C1" -> "C7" [label=" args[0]"]
  "C1" -> "C5" [label="info"]
  "C5" -> "C6" [label=" status"]
  "C7" -> "C8" [label=" strSize"]
  "C8" -> "C10" [label="strSize",constraint=false]
  "C10" -> "C11" [label=" myStr",constraint=false]
  "C11" -> "C12" [label=" myStr"]
  "C12" -> "C13" [label=" result"]
  "JS5" -> "C11" [style=invis]
}};
    \draw[red,ultra thick, dashed] (11.2,13.15) ellipse (1.4cm and 1cm);
\end{tikzpicture}}
\vspace{-30pt}
\caption{A cross-language data-flow graph for our example shown in Figure~\ref{fig:example}. We depict the JavaScript nodes with blue and the C/C++ ones with green. {Numbers denote line numbers in Figure~\ref{fig:example}.} 
With red we mark a potential location for sanitization in the JavaScript front-end.
}
\label{fig:example_dot}
\end{figure}

%% file: figs/type_conv.tex
    \pgfplotstableread{
Label Typecheck Suspicious Exploited Falsepos CheckC topper
\texttt{*.ToLocal\\Checked()} 27 126 17 4 162 0.001
{Cast to Buffer} 39 97 18 13 226 0.001
{Cast to Function} 11 13 1 0 16 0.001
    }\testdata
    \begin{figure}
        \centering
        \small
        \begin{tikzpicture}[yscale=0.9]
    \begin{axis}[
        height=6.7 cm,
        ybar stacked,
        ymin=0,
        ymax=230,
        xtick=data,
        bar shift=-5,
        enlarge x limits=0.4,
        legend style={
            at={(1,1.27)},
            draw=none,
            legend columns=2
        },
        reverse legend=true,
        xticklabels from table={\testdata}{Label},
        xticklabel style={text width=1.7cm,align=center},
        ylabel={Number of packages},
        xlabel={Type conversion APIs},
        ylabel style={yshift=-0.2cm},
        xlabel style={yshift=-0.4cm},
    ]
    \addplot [fill=solablue!40] table [y=Typecheck, meta=Label, x expr=\coordindex] {\testdata};
    \addlegendentry{Type check in JavaScript}
    \addplot [fill=red!40] table [y=Falsepos, meta=Label, x expr=\coordindex] {\testdata};
    \addlegendentry{False positives}
    \addplot [fill=solaorange!40] table [y=Suspicious, meta=Label, x expr=\coordindex] {\testdata};
    \addlegendentry{Unable to verify}
    \addplot [fill=ggreen,
    point meta=y] table [y=Exploited, meta=Label, x expr=\coordindex] {\testdata};
    \addlegendentry{Exploited}
    \end{axis}
    
    \begin{axis}[hide axis, bar shift=7pt,
    ybar stacked,
        ymin=0,
        ymax=230,
        xtick=data,
        enlarge x limits=0.4,
        legend style={
            at={(0.8,1.11)},
            draw=none,
            legend columns=2
        },
        height=6.7 cm
        ]
    \addplot[fill=gray!50] table [y = CheckC, meta=Label, x expr=\coordindex] {\testdata};
    \addlegendentry{Type check in C/C++}
\end{axis}
    
    \end{tikzpicture}
    \caption{Packages with flows to  type conversion APIs. The first bar represents unsanitized flows while the second one depicts sanitized flows.
    We further split unsanitized flows in different categories based on our manual inspection.}
        \label{fig:typecheck}
    \vspace{-15pt}
    \end{figure}
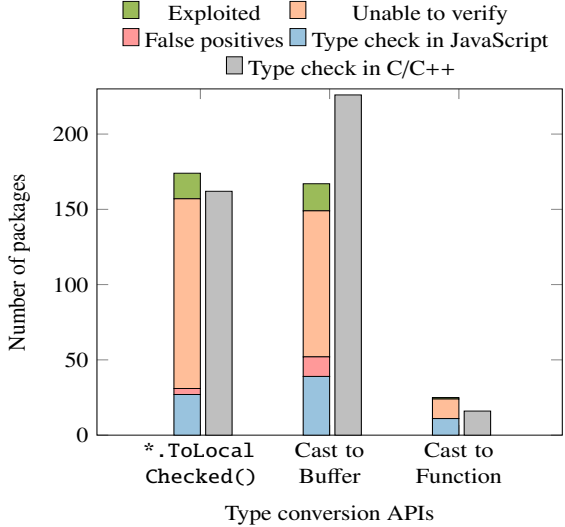

%% file: related-work.tex
\section{Related Work}
\label{sec:relwork}

\textbf{Binding layer and engine issues.} Vulnerabilities in JavaScript engines and in binding layer code seriously undermine the security guarantees of the language~\cite{BrownNWEJS17, ParkDGNVF20}. Analyzing this code got a lot of traction recently~\cite{BrownNWEJS17, dinh2021favocado, BrownSE20, HanOC19, Wang0WL19, HollerHZ12, ParkXYJK20, LeeHCS20, ParkDGNVF20}. The work in this domain can be categorized in two groups: fuzzing-based~\cite{HollerHZ12, funfuzz, HanOC19, LeeHCS20, ParkXYJK20, dinh2021favocado} and static analysis-based ~\cite{BrownNWEJS17, BrownSE20} approaches. 
Holler~{et al.} proposed LangFuzz~\cite{HollerHZ12}, which found 105 severe vulnerabilities in Mozilla's JavaScript interpreter. Given a set of seed programs, LangFuzz generates test cases 
by combining fragments of the seed programs. 
Instead of seed programs, Mozilla Security's FunFuzz~\cite{funfuzz} generates test cases from context-free grammars. The main limitation of these 
solutions  is the lack of semantic-awareness. Han~{et al.}~\cite{HanOC19} fixes this problem by proposing a code combining mechanism that uses a def-use analysis to find snippets with important semantic dependencies. Favocado~\cite{dinh2021favocado} is the first fuzzing-based tool to detect binding layer bugs. Favocado extracts semantic information from the API references and uses this to generate semantic-aware test cases. Sys~\cite{BrownSE20} is an analysis framework that combines static analysis and selective symbolic execution to identify low-level vulnerabilities in browser code. 
The most closely related work to ours is Brown~{et al.'s}~\cite{BrownNWEJS17} approach to find binding layer issues in JavaScript runtimes. Specifically, they describe various  bugs that undermine crash-, type- and memory-safety of the scripting language and propose using lightweight static checkers written in $\mu$chex~\cite{DBLP:conf/asplos/BrownNE16}. By using these checkers, they detect high profile vulnerabilities in the analyzed runtimes, showing the severity of the problem. 
In this work we study native extensions, which democratize the access to low-level code to non-expert users. Our results confirm that many of the issues introduced by Brown~{et al.'s}~\cite{BrownNWEJS17} are also prevalent in this new setting.
However, Brown et al.~\cite{BrownNWEJS17} use a much stronger attacker model that assumes that JavaScript code is untrusted, hence there is no need for cross-language analysis. On the contrary, we assume that the JavaScript part is benign, but vulnerable. We also consider languages beyond JavaScript to understand how API design decisions can enable misuses.

\textbf{Unsafe APIs uses.}  Almanee~{et al.}~\cite{almaneetoo} show that developers have an inertia to update vulnerable native libraries in Android apps, which consequently makes these apps vulnerable. Zimmermann~{et al.}~\cite{ZimmermannSTP19} show that the problem is prevalent in the Node.js ecosystem as well. Mastrangelo~{et al.}~\cite{MastrangeloPMLH15} show that  third-party library developers use unsafe Java virtual machine APIs for the sake of performance, which seriously undermines the security guarantees provided by the language. Evans~{et al.}~\cite{EvansCS20} show that the use of unsafe Rust features is widespread as well. To minimize the impact of unsafe Rust, Liu~{et al.} propose XRust~\cite{EvansCS20}, which ensures data integrity by logically dividing the safe and unsafe memory allocations into two mutually exclusive regions. Studies showed that there is a widespread tendency to misuse non-native APIs as well. For example, Java developers often misuse common platform-provided library APIs, e.g., Crypto APIs~\cite{RahamanXASTFKY19,EgeleBFK13,Acar0FGKMS17,KrugerS0BM18}, SSL/TLS APIs~\cite{FahlHMSBF12}, Fingerprint APIS~\cite{BianchiFMKVCL18}, as well as non-system APIs~\cite{DBLP:conf/sp/ZuoLZ19}.

\textbf{Node.js security.} Analyzing the security of the Node.js ecosystem has been a very active research field recently. Related work studies several threats in this ecosystem: regular expression denial-of-service~\cite{StaicuP18,DavisCSL18, DavisWL18, davisusing,Davis19}, code injections~\cite{StaicuPL18,GauthierHJ18,KarimTSS2018}, path traversals~\cite{LiangThesis2018}, trivial packages~\cite{AbdalkareemNWMS17,Kula2017}, hidden property abuse~\cite{xiao2021abusing}, vulnerable dependencies~\cite{DecanMC18}, APIs~\cite{DBLP:conf/sp/TalyEMMN11} and supply chain attacks~\cite{ZimmermannSTP19,duantowards}. While not strictly Node.js-specific, the adoption of WebAssembly may also pose additional risks for the runtime~\cite{0002KP20}. Existing solutions for reducing the attack surface of web applications using third-party code include package vetting~\cite{StaicuPL18,duantowards}, compartmentalization~\cite{VasilakisKRDDS18}, and debloating~\cite{koishybayev2020mininode}. We are the first to study the  native extensions' risks in this context.

\textbf{Comparative analysis of scripting languages.} Related work studies various security issues across multiple languages. Decan~et~al.~\cite{DecanMC17} and Kikas et al.~\cite{KikasGDP17} were the first to analyze the structure of third-party dependencies in various programming languages, and draw conclusions about interesting trends. More recently, Duan et al.~\cite{duantowards} proposes a technique for detecting supply chain attacks for the same set of scripting languages we consider in our work. {As discussed in Section~\ref{sec:discussion}, by integrating the same data flow analysis tools used by Duan et al.~\cite{duantowards} and by  modelling additional sinks and sources, our prototype can be extended to support Python and Ruby as well.}
Nonetheless, we are the first to perform an in-depth security study of an equivalent API in different scripting languages.

\textbf{Cross-language program analysis.} Researchers study various static analysis approaches to augment the insights of non-Java code to detect cross-lingual vulnerabilities in Java and Android applications~\cite{TanM07, TanC08, LeeDR16, ArztKB16, BaeLR19, LeeLR20,BruckerH16}. Nguyen \textit{et al.} built a cross-language program slicing framework to analyze PHP, HTML and JavaScript code in the same context~\cite{NguyenKN15}. Brucker \textit{et al.} propose static cross-language call graphs for hybrid JavaScript/Java mobile apps. There has been attempts to build cross-language dynamic taint analysis platforms as well~\cite{KreindlBM19, KreindlBSLM20,BaiWQZWP19}. Alimadadi \textit{et al.} propose an approach to model temporal and behavioral information in full-stack JavaScript applications to detect cross-stack bugs~\cite{Alimadadi0P16}. 
Recently, Hu \textit{et al.}~\cite{HuZ20} study the usage of native extension API in Python, while Monat \textit{et al.}~\cite{MonatOM21} propose a sophisticated static analysis for detecting cross-language bugs in Python.
We are the first to perform cross-language analysis in the context of native extensions security. 

%% file: figs/bar_types.tex
\begin{figure}
    \centering
    \small
\begin{tikzpicture}
  \begin{axis}[
  height=4cm,
  width=\linewidth,
    axis x line*=bottom,
    axis y line*=none,
    every outer y axis line/.append style={draw=none},
    every y tick/.append style={draw=none},
    ymin=0,
    ymax=3500,
    symbolic x coords={Object, String, Boolean, Number, Function, Array/Buffer},
    ytick={0, 1000, 2000, 3000, 4000},
    x tick label style={rotate=45,anchor=east},
    ymajorgrids,
    y grid style={densely dotted, line cap=round},
    ylabel={Number of packages},
    xlabel style={yshift=-0.8cm},
    ylabel style={yshift=0cm},
    nodes near coords,
  ]
    \addplot[
      ybar,
      draw=black,
      fill=solablue!40
    ] coordinates {
        (Object, 3376)
        (String, 2740)
        (Boolean, 321)
        (Number, 3249)
        (Function, 1986)
        (Array/Buffer, 486)
    };
  \end{axis}
\end{tikzpicture}
\vspace{-5pt}
    \caption{Number of packages explicitly converting values to various C/C++ types. 
    \vspace{-15pt}
    }
    \label{fig:types_package}
\end{figure}
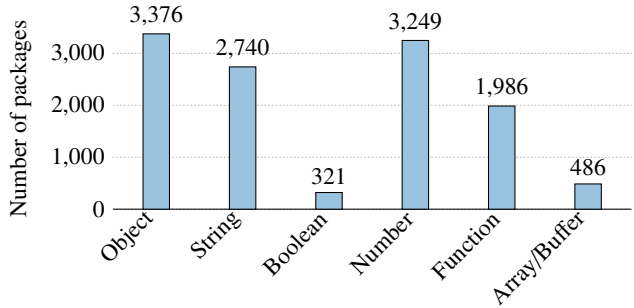

%% file: figs/appedix.tex
\begin{figure*}[ph!]
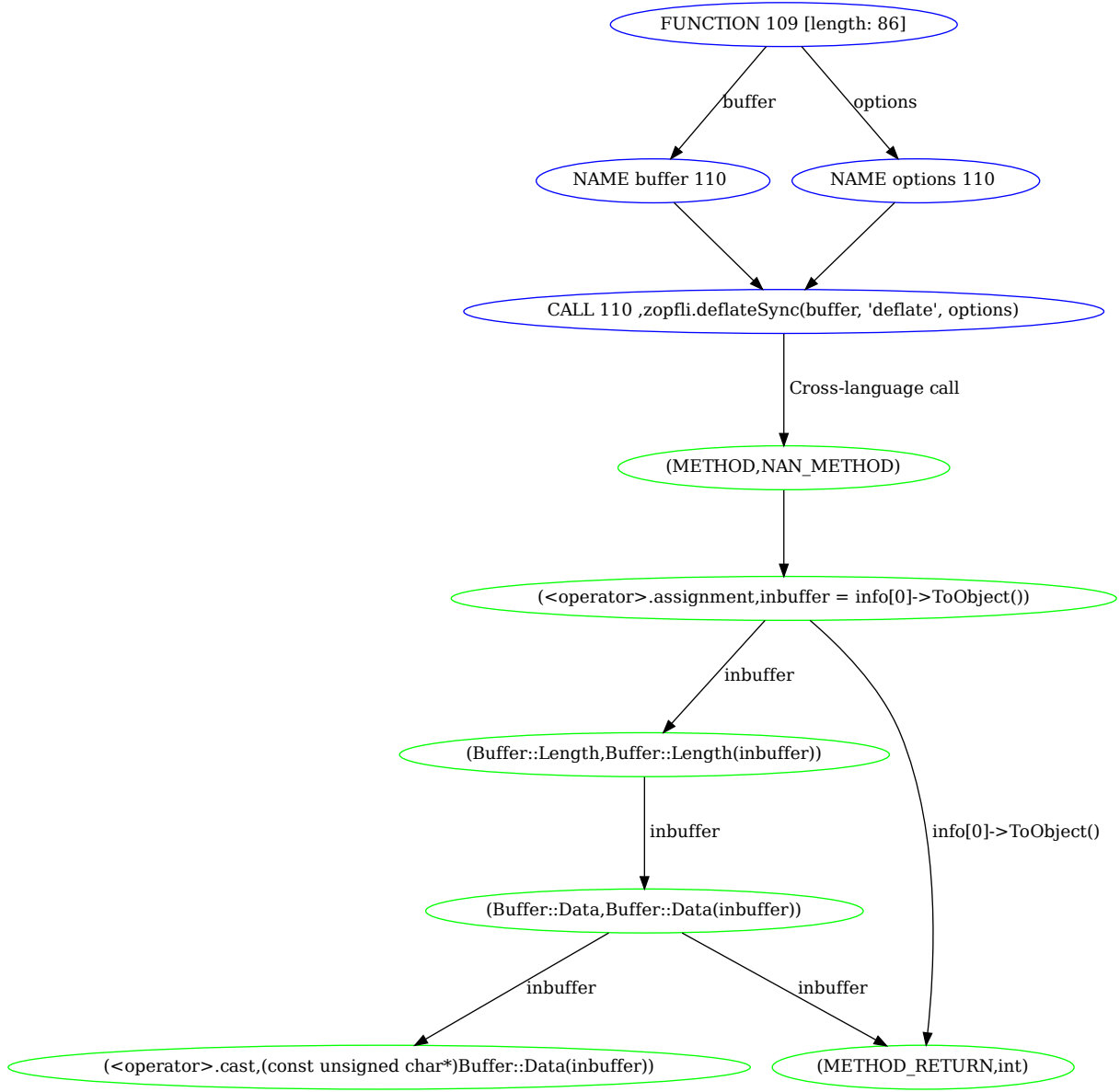

    \centering
    \resizebox{\linewidth}{!}{
\digraph{abc}{
ratio=1
  "0" [label="FUNCTION 109 [length: 86]",color=blue]
  "4" [label="NAME buffer 110 ",color=blue]
  "5" [label="NAME options 110 ",color=blue]
  "9" [label="CALL 110 ,zopfli.deflateSync(buffer, 'deflate', options)",color=blue]
  "249280325321372575" [label="(METHOD,NAN_METHOD)",color=green]
  "249280325321372585" [label="(<operator>.assignment,inbuffer = info[0]->ToObject())",color=green]
  "249280325321372591" [label="(Buffer::Length,Buffer::Length(inbuffer))",color=green]
  "249280325321372598" [label="(Buffer::Data,Buffer::Data(inbuffer))",color=green]
  "249280325321372620" [label="(METHOD_RETURN,int)",color=green]
  "249280325321372596" [label="(<operator>.cast,(const unsigned char*)Buffer::Data(inbuffer))",color=green]
  "0" -> "4" [label=buffer]
  "0" -> "5" [label=options]
  "4" -> "9" [label=""]
  "5" -> "9" [label=""]
  "9" -> "249280325321372575" [label=" Cross-language call"]
  "249280325321372575" -> "249280325321372585"
  "249280325321372585" -> "249280325321372620" [label=" info[0]->ToObject()"]
  "249280325321372585" -> "249280325321372591" [label=" inbuffer"]
  "249280325321372591" -> "249280325321372598" [label=" inbuffer"]
  "249280325321372598" -> "249280325321372620" [label=" inbuffer"]
  "249280325321372598" -> "249280325321372596" [label=" inbuffer"]
}
}
    \caption{A cross-language data-flow graph for the package \texttt{zopfli-node}, version 2.0.3. The graph corresponds to the method \texttt{CompressBinding::Sync} in C++ and \texttt{lib/zopfli.js} in JavaScript. With blue we show the JavaScript nodes and with green the C++ ones. The graph shows an unsanitized data flow from the \texttt{buffer} parameter in JavaScript to the type conversion API \texttt{Buffer::Data(*)} in C++.}
    \label{fig:zopfli}
\end{figure*}

\begin{figure*}[ph!]
    \centering
    \resizebox{\linewidth}{!}{
\digraph{iltorb}{
ratio=1
  "0" [label="FUNCTION compressSync 43 [length: 201]",color=blue]
  "4" [label="NAME input 44 ",color=blue]
  "5" [label="NAME params 47 ",color=blue]
  "6" [label="NAME params 47 ",color=blue]
  "7" [label="NAME input 48 ",color=blue]
  "8" [label="NAME params 48 ",color=blue]
  "12" [label="EXPR_RESULT 47 ,params = params || {}",color=blue]
  "14" [label="CALL 44 ,Buffer.isBuffer(input)",color=red]
  "15" [label="OR 47 ",color=blue]
  "16" [label="ASSIGN 47 ,params = params || {}",color=blue]
  "17" [label="CALL 48 ,encode.compressSync(input, params)",color=blue]
  "18" [label="NOT 44 ",color=blue]
  "4534824584787439935" [label="(METHOD,NAN_METHOD)",color=green]
  "4534824584787439944" [label="(<operator>.assignment,buffer = info[0]->ToObject())",color=green]
  "4534824584787439949" [label="(node::Buffer::Length,node::Buffer::Length(buffer))",color=green]
  "4534824584787439953" [label="(node::Buffer::Data,node::Buffer::Data(buffer))",color=green]
  "4534824584787439970" [label="(METHOD_RETURN,int)",color=green]
  "4534824584787439948" [label=<(compressor.CopyInputToRingBuffer,<BR/>compressor.CopyInputToRingBuffer(node::Buffer::Length(buffer), <BR/> (uint8_t*) node::Buffer::Data(buffer)))>,color=green]
  "4534824584787439951" [label="(<operator>.cast,(uint8_t*) node::Buffer::Data(buffer))",color=green]
  "0" -> "4" [label=input]
  "0" -> "5" [label=params]
  "0" -> "6" [label=params]
  "0" -> "7" [label=input]
  "0" -> "8" [label=params]
  "4" -> "14" [label=""]
  "5" -> "15" [label=""]
  "6" -> "16" [label=""]
  "7" -> "17" [label=""]
  "8" -> "17" [label=""]
  "14" -> "18" [label=""]
  "15" -> "16" [label=""]
  "16" -> "12" [label=""]
  "17" -> "4534824584787439935" [label=" Cross-language call"]
  "4534824584787439935" -> "4534824584787439944"
  "4534824584787439944" -> "4534824584787439970" [label=" info[0]->ToObject()"]
  "4534824584787439944" -> "4534824584787439949" [label=" buffer"]
  "4534824584787439949" -> "4534824584787439948" [label=" buffer"]
  "4534824584787439949" -> "4534824584787439953" [label=" buffer"]
  "4534824584787439953" -> "4534824584787439970" [label=" buffer"]
  "4534824584787439953" -> "4534824584787439951" [label=" buffer"]
  "4534824584787439951" -> "4534824584787439948" [style=invis]
  "4534824584787439951" -> "4534824584787439970" [style=invis]
}
}
    \caption{A cross-language data-flow graph for the package \texttt{iltorb}, version 1.0.0. The graph corresponds to the method \texttt{compressSync} in C++ and \texttt{index.js} in JavaScript. With blue we show the JavaScript nodes and with green the C++ ones. The graph shows a sanitized data flow from the \texttt{input} parameter in JavaScript to the type conversion API \texttt{\texttt{Buffer::Data(*)}} in C++. The sanitizer is marked with red and it can be found in the JavaScript part of the code.}
    \label{fig:iltorb}
\end{figure*}

\begin{figure*}[ph!]
    \centering
    \resizebox{\linewidth}{!}{
\digraph{bcrypt}{
ratio=1.1
node [nodesep=0.5, fontsize=28];
edge [ fontsize=22, lblstyle="above, sloped" ];
  "0" [label="FUNCTION 8 [length: 256]",color=blue]
  "4" [label="NAME rounds 10 ",color=blue]
  "7" [label="NAME rounds 16 ",color=blue]
  "5" [label="NAME rounds 11 ",color=blue]
  "6" [label="NAME rounds 12 ",color=blue]
  "11" [label="NOT 10 ",color=blue]
  "12" [label="ASSIGN 11 ,rounds = 10",color=blue]
  "13" [label="TYPEOF 12 ",color=red]
  "14" [label=<CALL 16 ,<BR/>bindings.gen_salt_sync(rounds,<BR/> crypto.randomBytes(16))>,color=blue]
  "19" [label="EXPR_RESULT 11 ,rounds = 10",color=blue]
  "20" [label="SHNE 12 ",color=blue]
  "23" [label="IF 12 ",color=blue]
  "1152921504607722372" [label="(METHOD,NAN_METHOD)",color=green]
  "1152921504607722383" [label=<(&lt;operator&gt;.notEquals,<BR/>Buffer::Length(args[1]) != 16)>,color=green]
  "1152921504607722396" [label=<(&lt;operator&gt;.assignment,<BR/>rounds = args[0]-&gt;Int32Value())>,color=green]
  "1152921504607722400" [label=<(&lt;operator&gt;.assignment,<BR/>* seed = (u_int8_t*)Buffer::Data(args[1]))>,color=green]
  "1152921504607722409" [label=<(bcrypt_gensalt,<BR/> bcrypt_gensalt(rounds, seed, salt))>,color=green] 
  "1152921504607722416" [label="(strlen,strlen(salt))",color=green]
  "1152921504607722419" [label="(METHOD_RETURN,int)",color=green]
  "1152921504607722377" [label=<(&lt;operator&gt;.logicalOr,<BR/>!Buffer::HasInstance(args[1]) || <BR/> Buffer::Length(args[1]) != 16)>,color=red]
  "1152921504607722398" [label=<(args[0]-&gt;Int32Value,<BR/>args[0]-&gt;Int32Value())>,color=green]
  "1152921504607722405" [label=<(&lt;operator&gt;.indirectIndexAccess,<BR/>args[1])>,color=green]
  "1152921504607722402" [label=<(&lt;operator&gt;.cast,<BR/>(u_int8_t*)Buffer::Data(args[1]))>,color=green]
  "1152921504607722404" [label=<(Buffer::Data,<BR/>Buffer::Data(args[1]))>,color=green]
  "0" -> "4" [label=rounds]
  "0" -> "5" [label=rounds]
  "0" -> "6" [label=rounds]
  "0" -> "7" [label=rounds]
  "4" -> "11" [label=""]
  "5" -> "12" [label=""]
  "6" -> "13" [label=""]
  "7" -> "14" [label=""]
  "12" -> "19" [label=""]
  "13" -> "20" [label=""]
  "14" -> "1152921504607722372" [label=" Cross-language call"]
  "20" -> "23" [label=""]
    "1152921504607722372" -> "1152921504607722383"
  "1152921504607722372" -> "1152921504607722396"
  "1152921504607722372" -> "1152921504607722400"
  "1152921504607722383" -> "1152921504607722377" [label=" 16"]
  "1152921504607722383" -> "1152921504607722419" [label=" Buffer::Length(args[1])"]
  "1152921504607722396" -> "1152921504607722419" [label=" args[0]->Int32Value()"]
  "1152921504607722396" -> "1152921504607722409" [label=" rounds"]
  "1152921504607722400" -> "1152921504607722419" [label=" (u_int8_t*)Buffer::Data(args[1])"]
  "1152921504607722400" -> "1152921504607722409" [label=" seed"]
  "1152921504607722409" -> "1152921504607722419" [label=" seed"]
  "1152921504607722409" -> "1152921504607722416" [label=" salt"]
  "1152921504607722377" -> "1152921504607722398"
  "1152921504607722377" -> "1152921504607722405"
  "1152921504607722377" -> "1152921504607722402"
  "1152921504607722377" -> "1152921504607722404"
  "1152921504607722405" -> "1152921504607722404"
  "1152921504607722400" -> "1152921504607722396" [style=invis]
  "1152921504607722402"  -> "1152921504607722398" [style=invis]
  "1152921504607722377" -> "1152921504607722396" [style=invis]
  "1152921504607722402" -> "1152921504607722398" [style=invis]
}
}
    \caption{A cross-language data-flow graph for the package \texttt{nan-bcrypt}, version 0.7.7. The graph corresponds to the method \texttt{gen\_salt\_sync} in C++ and \texttt{bcrypt.js} in JavaScript. With blue we show the JavaScript nodes and with green the C++ ones. The graph shows a sanitized data flow from the \texttt{rounds} parameter in JavaScript to the type conversion API \texttt{Int32Value()} in C++. The sanitizers are marked with red and can be found both in the JavaScript and in the C++ part of the code.}
    \label{fig:bcrypt}
\end{figure*}

%% file: perf.tex
{\centering
\begin{tikzpicture}
\begin{axis}[
    clip=false,
    ymin=0,ymax=1,
    xmin=0, xmax=903,
    ylabel={Ratio of packages},
    xlabel={Time (seconds)},
    xtick = {0,300,600,900},
    every axis plot/.style={line width=2pt},
    legend style={
            at={(0.95,0.25)},
            },
    table/create on use/cumulative distribution/.style={
        create col/expr={\pgfmathaccuma + \thisrow{f(x)}}   
    }
]
\addplot+ [mark=none, blue] table [y=cumulative distribution]{
x f(x)
1 0
2 0.0941486
3 0.0973044
4 0.120184
5 0.0895464
6 0.069954
7 0.12071
8 0.0786325
9 0.0483892
10 0.034977
11 0.0245891
12 0.0230112
13 0.0135437
14 0.0127548
15 0.00959895
16 0.0109139
17 0.00841552
18 0.00591716
19 0.0052597
20 0.00552268
21 0.0034188
22 0.0035503
23 0.0052597
24 0.00499671
25 0.00302433
26 0.00315582
27 0.0034188
28 0.00262985
29 0.00184089
30 0.00262985
31 0.00197239
32 0.00197239
33 0.00236686
34 0.00184089
35 0.000920447
36 0.00197239
37 0.00118343
38 0.00118343
39 0.00184089
40 0.00144642
41 0.00249836
42 0.00144642
43 0.00157791
44 0.000920447
45 0.00144642
46 0.00157791
47 0.00105194
48 0.00105194
49 0.000788955
50 0.00105194
51 0.0017094
52 0.000657462
53 0.00052597
54 0.000920447
55 0.000131492
56 0.000657462
57 0.00118343
58 0.000394477
59 0.000262985
60 0.000657462
61 0.000788955
62 0.000788955
63 0.000788955
64 0.000657462
65 0.000394477
66 0.000657462
67 0.00052597
68 0.00052597
69 0.000788955
70 0.00052597
71 0.000920447
72 0.000657462
73 0.000394477
74 0.000657462
75 0.00052597
76 0.00052597
77 0.000262985
78 0.000262985
79 0.000131492
80 0.000262985
81 0.000394477
82 0.00118343
83 0.000262985
84 0.000394477
85 0.000394477
86 0.000394477
87 0.000657462
88 0.00052597
89 0.000131492
90 0.000131492
91 0.000657462
92 0.000394477
93 0.000262985
94 0.000131492
95 0.000262985
96 0.00052597
97 0.000131492
98 0.000131492
99 0.000262985
100 0.000131492
101 0.000131492
102 0.000262985
103 0.000131492
104 0.000657462
105 0.000262985
106 0.000131492
107 0.000131492
108 0.000262985
109 0.000131492
110 0.000394477
111 0
112 0.000262985
113 0.00052597
114 0.000262985
115 0.00052597
116 0.000131492
117 0.000394477
118 0.000262985
119 0.000788955
120 0.000394477
121 0
122 0.000131492
123 0.000394477
124 0.000262985
125 0
126 0
127 0
128 0.000131492
129 0
130 0
131 0.000262985
132 0.000394477
133 0
134 0.000262985
135 0
136 0.000131492
137 0.000394477
138 0.000131492
139 0.000262985
140 0.000657462
141 0.000262985
142 0.000394477
143 0.000131492
144 0
145 0.000262985
146 0.000262985
147 0
148 0.000262985
149 0
150 0.000262985
151 0.000131492
152 0.000131492
153 0
154 0.000131492
155 0
156 0.00052597
157 0.000131492
158 0
159 0.000131492
160 0
161 0.000131492
162 0
163 0.000131492
164 0
165 0.000262985
166 0.000131492
167 0.000262985
168 0.000131492
169 0
170 0
171 0.000131492
172 0.000262985
173 0
174 0
175 0.000131492
176 0
177 0.000131492
178 0
179 0.000131492
180 0
181 0.000131492
182 0
183 0
184 0.000131492
185 0
186 0
187 0
188 0.000131492
189 0.000131492
190 0
191 0
192 0
193 0.000131492
194 0
195 0
196 0.000131492
197 0
198 0.000131492
199 0
200 0.000262985
201 0
202 0
203 0.000131492
204 0
205 0.000131492
206 0
207 0.000262985
208 0.000131492
209 0.000131492
210 0
211 0.000131492
212 0
213 0
214 0
215 0.000262985
216 0
217 0.000131492
218 0
219 0.000262985
220 0
221 0
222 0.000131492
223 0
224 0
225 0
226 0
227 0.000262985
228 0.000131492
229 0.000131492
230 0
231 0.000131492
232 0.000131492
233 0.000262985
234 0
235 0.000394477
236 0.000131492
237 0
238 0.000131492
239 0
240 0
241 0.000131492
242 0
243 0
244 0
245 0
246 0
247 0
248 0
249 0
250 0
251 0.000131492
252 0
253 0
254 0
255 0.000131492
256 0.000131492
257 0
258 0.000131492
259 0
260 0.000131492
261 0
262 0
263 0
264 0
265 0
266 0.000262985
267 0
268 0
269 0.000131492
270 0.000131492
271 0
272 0
273 0.000131492
274 0
275 0
276 0
277 0.000131492
278 0
279 0.000131492
280 0
281 0
282 0
283 0
284 0
285 0
286 0
287 0
288 0
289 0
290 0
291 0.000131492
292 0
293 0
294 0
295 0
296 0.000262985
297 0
298 0
299 0
300 0
301 0
302 0
303 0
304 0
305 0
306 0
307 0
308 0
309 0
310 0
311 0
312 0
313 0
314 0
315 0
316 0.000131492
317 0
318 0
319 0
320 0
321 0
322 0
323 0.000131492
324 0
325 0.000131492
326 0
327 0.000131492
328 0.000131492
329 0
330 0
331 0.000131492
332 0
333 0
334 0
335 0
336 0
337 0.000131492
338 0
339 0
340 0.000131492
341 0
342 0
343 0.000131492
344 0
345 0
346 0
347 0
348 0
349 0
350 0
351 0
352 0
353 0
354 0
355 0
356 0
357 0.000131492
358 0
359 0
360 0
361 0
362 0
363 0
364 0
365 0.000131492
366 0
367 0
368 0
369 0
370 0
371 0
372 0
373 0
374 0.000131492
375 0
376 0
377 0
378 0
379 0
380 0.000131492
381 0
382 0.000262985
383 0
384 0
385 0
386 0.000131492
387 0.000131492
388 0
389 0
390 0
391 0
392 0
393 0
394 0
395 0
396 0
397 0
398 0
399 0.000131492
400 0
401 0
402 0
403 0
404 0
405 0
406 0
407 0
408 0
409 0
410 0
411 0.000131492
412 0
413 0.000262985
414 0
415 0
416 0.000131492
417 0
418 0
419 0
420 0
421 0
422 0
423 0
424 0
425 0
426 0
427 0
428 0
429 0
430 0
431 0
432 0
433 0
434 0
435 0
436 0
437 0
438 0
439 0
440 0
441 0
442 0
443 0
444 0
445 0
446 0
447 0
448 0
449 0
450 0
451 0
452 0
453 0
454 0
455 0
456 0
457 0
458 0
459 0
460 0
461 0
462 0
463 0
464 0
465 0
466 0
467 0.000262985
468 0
469 0
470 0
471 0
472 0
473 0
474 0.000131492
475 0
476 0
477 0
478 0
479 0
480 0
481 0
482 0
483 0
484 0
485 0
486 0
487 0
488 0
489 0
490 0
491 0
492 0
493 0
494 0
495 0
496 0
497 0.000131492
498 0
499 0
500 0
501 0
502 0
503 0
504 0
505 0
506 0
507 0
508 0
509 0
510 0
511 0
512 0.000131492
513 0
514 0
515 0
516 0
517 0
518 0
519 0
520 0
521 0
522 0
523 0
524 0
525 0.000131492
526 0
527 0
528 0
529 0
530 0
531 0
532 0
533 0
534 0
535 0
536 0
537 0
538 0
539 0
540 0
541 0
542 0.000131492
543 0
544 0
545 0
546 0.000131492
547 0
548 0.000131492
549 0
550 0
551 0
552 0
553 0
554 0
555 0
556 0
557 0
558 0.000131492
559 0
560 0
561 0
562 0
563 0.000131492
564 0
565 0
566 0
567 0
568 0.000131492
569 0
570 0
571 0
572 0.000131492
573 0
574 0
575 0
576 0
577 0
578 0
579 0
580 0
581 0
582 0.000131492
583 0
584 0
585 0
586 0.000131492
587 0
588 0
589 0
590 0.000131492
591 0
592 0
593 0
594 0
595 0
596 0
597 0
598 0
599 0
600 0
601 0
602 0
603 0.000131492
604 0.000131492
605 0
606 0.000131492
607 0
608 0.000131492
609 0
610 0.000131492
611 0
612 0
613 0
614 0
615 0
616 0
617 0
618 0
619 0
620 0
621 0
622 0
623 0
624 0
625 0
626 0
627 0
628 0
629 0
630 0
631 0.000131492
632 0
633 0
634 0
635 0
636 0
637 0
638 0
639 0
640 0
641 0
642 0
643 0
644 0
645 0
646 0
647 0
648 0.000131492
649 0
650 0
651 0
652 0
653 0
654 0
655 0
656 0
657 0
658 0
659 0
660 0
661 0
662 0
663 0
664 0
665 0
666 0
667 0
668 0
669 0
670 0
671 0
672 0
673 0
674 0
675 0
676 0
677 0
678 0
679 0
680 0
681 0
682 0
683 0
684 0
685 0
686 0
687 0
688 0
689 0
690 0.000131492
691 0
692 0
693 0
694 0
695 0
696 0
697 0
698 0
699 0
700 0
701 0
702 0
703 0
704 0
705 0
706 0
707 0
708 0
709 0
710 0
711 0
712 0
713 0
714 0
715 0
716 0
717 0
718 0
719 0
720 0.000131492
721 0
722 0
723 0
724 0
725 0
726 0.000131492
727 0
728 0
729 0
730 0
731 0
732 0
733 0
734 0
735 0
736 0
737 0
738 0
739 0
740 0
741 0
742 0
743 0
744 0
745 0
746 0
747 0.000131492
748 0
749 0
750 0
751 0
752 0
753 0
754 0
755 0
756 0
757 0
758 0
759 0
760 0
761 0
762 0
763 0
764 0
765 0
766 0
767 0
768 0
769 0
770 0
771 0
772 0
773 0
774 0
775 0
776 0
777 0
778 0
779 0
780 0
781 0
782 0
783 0.000131492
784 0
785 0
786 0
787 0
788 0
789 0
790 0
791 0
792 0
793 0
794 0
795 0
796 0
797 0
798 0
799 0
800 0
801 0
802 0
803 0
804 0
805 0
806 0
807 0
808 0
809 0
810 0
811 0
812 0
813 0
814 0
815 0
816 0
817 0
818 0
819 0
820 0
821 0.000131492
822 0
823 0
824 0
825 0
826 0
827 0
828 0
829 0
830 0
831 0
832 0
833 0
834 0
835 0
836 0.000131492
837 0
838 0
839 0
840 0
841 0
842 0
843 0
844 0
845 0
846 0
847 0
848 0
849 0.000131492
850 0
851 0
852 0
853 0
854 0
855 0
856 0
857 0
858 0
859 0
860 0
861 0
862 0
863 0
864 0
865 0
866 0
867 0
868 0
869 0
870 0
871 0
872 0
873 0
874 0
875 0
876 0
877 0
878 0
879 0
880 0
881 0
882 0
883 0
884 0.000131492
885 0
886 0
887 0
888 0
889 0
890 0
891 0
892 0
893 0
894 0
895 0
896 0
897 0
898 0
899 0
900 0
901 0.00894149
902 0
903 0
};
\addlegendentry{C/C++}
\addplot+ [mark=none, green] table [y=cumulative distribution]{
x f(x)
1 0.269691
2 0.463905
3 0.110454
4 0.0403682
5 0.0216963
6 0.0142012
7 0.00749507
8 0.00578567
9 0.00276134
10 0.0034188
11 0.0035503
12 0.0035503
13 0.00197239
14 0.00184089
15 0.0017094
16 0.00105194
17 0.000920447
18 0.00184089
19 0.00197239
20 0.00144642
21 0.000788955
22 0.00118343
23 0.000788955
24 0.000788955
25 0.000657462
26 0.00118343
27 0.00052597
28 0.00052597
29 0.000394477
30 0.00052597
31 0.000262985
32 0.000131492
33 0.000131492
34 0.000788955
35 0.000131492
36 0
37 0
38 0.000394477
39 0.000657462
40 0.000394477
41 0
42 0.000657462
43 0.000394477
44 0.00052597
45 0.000262985
46 0.000394477
47 0.000394477
48 0.000131492
49 0.000131492
50 0.000131492
51 0
52 0.000131492
53 0.000131492
54 0.000131492
55 0.000131492
56 0.000262985
57 0.000788955
58 0.000262985
59 0.000262985
60 0
61 0.000657462
62 0.000394477
63 0.000131492
64 0.000262985
65 0.00052597
66 0.000394477
67 0.000262985
68 0.000262985
69 0
70 0.000131492
71 0.000131492
72 0
73 0
74 0.000262985
75 0
76 0.000131492
77 0
78 0.000262985
79 0.000131492
80 0.000131492
81 0.000131492
82 0.000131492
83 0
84 0.000394477
85 0.000131492
86 0
87 0.000262985
88 0.000131492
89 0.000131492
90 0
91 0
92 0
93 0
94 0
95 0
96 0.000131492
97 0
98 0.000262985
99 0
100 0.000131492
101 0.000131492
102 0
103 0
104 0
105 0
106 0
107 0
108 0.000131492
109 0.000131492
110 0
111 0
112 0
113 0
114 0
115 0.000131492
116 0
117 0
118 0.000131492
119 0
120 0
121 0
122 0
123 0
124 0.000131492
125 0
126 0
127 0
128 0
129 0.000131492
130 0
131 0
132 0
133 0
134 0
135 0
136 0
137 0.000131492
138 0
139 0
140 0.000131492
141 0
142 0.000131492
143 0
144 0
145 0
146 0.000262985
147 0
148 0
149 0.000131492
150 0.000262985
151 0
152 0.000394477
153 0.000131492
154 0.000262985
155 0.000131492
156 0.000131492
157 0.000394477
158 0.000262985
159 0.000394477
160 0.000131492
161 0.000131492
162 0
163 0.000131492
164 0.000262985
165 0.000131492
166 0.000131492
167 0.000131492
168 0
169 0
170 0
171 0.000262985
172 0.000131492
173 0
174 0
175 0
176 0
177 0
178 0
179 0
180 0
181 0
182 0
183 0
184 0
185 0
186 0
187 0
188 0
189 0
190 0
191 0
192 0
193 0
194 0
195 0
196 0
197 0
198 0
199 0
200 0
201 0
202 0
203 0
204 0
205 0
206 0
207 0.000131492
208 0
209 0
210 0.000131492
211 0
212 0
213 0
214 0.000131492
215 0.000131492
216 0
217 0
218 0.000262985
219 0.000262985
220 0.000262985
221 0
222 0
223 0.000262985
224 0
225 0.000131492
226 0
227 0
228 0.000131492
229 0.000131492
230 0
231 0
232 0
233 0.000131492
234 0
235 0
236 0
237 0.000131492
238 0
239 0
240 0
241 0.000131492
242 0
243 0.000131492
244 0.000131492
245 0
246 0
247 0
248 0
249 0
250 0
251 0
252 0
253 0
254 0
255 0
256 0
257 0
258 0
259 0
260 0
261 0
262 0
263 0
264 0
265 0
266 0
267 0
268 0
269 0
270 0
271 0.000131492
272 0
273 0
274 0.000131492
275 0
276 0
277 0
278 0
279 0.000131492
280 0
281 0
282 0
283 0
284 0
285 0
286 0
287 0
288 0
289 0
290 0
291 0
292 0
293 0
294 0.000131492
295 0
296 0
297 0.000131492
298 0
299 0
300 0
301 0.000131492
302 0
303 0.000131492
304 0.000131492
305 0
306 0
307 0
308 0
309 0
310 0
311 0
312 0
313 0
314 0
315 0.000131492
316 0
317 0
318 0
319 0
320 0
321 0
322 0
323 0
324 0
325 0
326 0
327 0
328 0
329 0
330 0
331 0
332 0
333 0.000131492
334 0
335 0
336 0
337 0
338 0
339 0
340 0
341 0
342 0
343 0
344 0.000131492
345 0.000131492
346 0
347 0
348 0.000131492
349 0
350 0
351 0
352 0
353 0
354 0.000131492
355 0
356 0
357 0.000131492
358 0
359 0
360 0
361 0
362 0
363 0
364 0
365 0
366 0
367 0
368 0
369 0
370 0
371 0
372 0
373 0
374 0
375 0
376 0
377 0
378 0
379 0
380 0
381 0
382 0
383 0
384 0
385 0
386 0
387 0
388 0
389 0
390 0.000131492
391 0
392 0
393 0
394 0
395 0
396 0
397 0
398 0
399 0
400 0
401 0
402 0
403 0
404 0
405 0
406 0.000131492
407 0
408 0
409 0.000131492
410 0
411 0
412 0
413 0
414 0
415 0
416 0
417 0
418 0
419 0
420 0
421 0
422 0
423 0
424 0
425 0.000131492
426 0
427 0
428 0
429 0
430 0
431 0
432 0
433 0
434 0.000131492
435 0
436 0
437 0
438 0
439 0.000131492
440 0
441 0.000131492
442 0
443 0
444 0
445 0
446 0
447 0
448 0
449 0
450 0
451 0
452 0
453 0
454 0
455 0
456 0.000262985
457 0
458 0
459 0
460 0
461 0
462 0.000131492
463 0
464 0
465 0
466 0.000131492
467 0
468 0
469 0
470 0
471 0
472 0
473 0
474 0.000131492
475 0
476 0
477 0
478 0.000131492
479 0
480 0
481 0.000131492
482 0
483 0.000131492
484 0
485 0
486 0
487 0
488 0
489 0
490 0
491 0
492 0
493 0
494 0
495 0
496 0
497 0
498 0.000131492
499 0
500 0
501 0.000131492
502 0.000131492
503 0
504 0
505 0
506 0
507 0
508 0
509 0
510 0
511 0
512 0
513 0
514 0
515 0
516 0
517 0
518 0
519 0.000131492
520 0
521 0
522 0.000131492
523 0
524 0.000131492
525 0
526 0
527 0
528 0
529 0
530 0.000131492
531 0
532 0
533 0
534 0
535 0
536 0
537 0
538 0
539 0
540 0
541 0
542 0
543 0
544 0
545 0
546 0
547 0
548 0
549 0
550 0
551 0
552 0
553 0
554 0.000131492
555 0
556 0
557 0
558 0
559 0
560 0
561 0
562 0
563 0.000131492
564 0
565 0
566 0
567 0
568 0
569 0
570 0
571 0
572 0
573 0
574 0
575 0
576 0.000131492
577 0
578 0
579 0
580 0
581 0
582 0
583 0
584 0
585 0
586 0
587 0
588 0
589 0
590 0.000131492
591 0
592 0
593 0
594 0
595 0
596 0
597 0
598 0
599 0
600 0
601 0
602 0
603 0
604 0
605 0
606 0
607 0
608 0
609 0
610 0
611 0
612 0
613 0
614 0.000131492
615 0
616 0
617 0
618 0
619 0
620 0
621 0
622 0
623 0
624 0
625 0
626 0
627 0
628 0
629 0
630 0
631 0
632 0
633 0
634 0
635 0
636 0
637 0
638 0
639 0
640 0
641 0
642 0
643 0
644 0
645 0
646 0.000131492
647 0
648 0
649 0
650 0
651 0
652 0
653 0
654 0
655 0
656 0
657 0
658 0
659 0
660 0
661 0
662 0
663 0
664 0
665 0
666 0
667 0
668 0
669 0
670 0
671 0
672 0
673 0
674 0
675 0
676 0
677 0
678 0
679 0
680 0
681 0
682 0
683 0
684 0
685 0
686 0
687 0
688 0
689 0
690 0
691 0
692 0
693 0
694 0
695 0
696 0
697 0
698 0
699 0
700 0
701 0
702 0
703 0
704 0
705 0
706 0
707 0
708 0
709 0
710 0
711 0
712 0
713 0
714 0
715 0
716 0
717 0
718 0
719 0
720 0
721 0
722 0
723 0
724 0
725 0
726 0
727 0.000131492
728 0
729 0
730 0
731 0
732 0
733 0
734 0
735 0
736 0
737 0
738 0
739 0
740 0
741 0
742 0
743 0
744 0
745 0
746 0
747 0
748 0
749 0
750 0
751 0
752 0
753 0
754 0
755 0
756 0
757 0.000131492
758 0
759 0.000131492
760 0
761 0
762 0
763 0
764 0
765 0
766 0
767 0
768 0
769 0
770 0
771 0
772 0
773 0
774 0
775 0
776 0
777 0
778 0
779 0
780 0
781 0
782 0
783 0
784 0
785 0
786 0
787 0
788 0
789 0
790 0
791 0
792 0
793 0
794 0
795 0
796 0
797 0
798 0
799 0
800 0
801 0
802 0
803 0
804 0
805 0
806 0
807 0
808 0
809 0
810 0
811 0
812 0
813 0
814 0
815 0
816 0
817 0
818 0
819 0
820 0
821 0
822 0
823 0
824 0
825 0
826 0
827 0
828 0
829 0
830 0
831 0
832 0
833 0
834 0
835 0
836 0
837 0
838 0
839 0
840 0
841 0
842 0
843 0
844 0
845 0
846 0
847 0.000131492
848 0
849 0
850 0
851 0
852 0
853 0
854 0
855 0
856 0.000131492
857 0
858 0
859 0
860 0
861 0
862 0
863 0.000131492
864 0
865 0
866 0
867 0
868 0
869 0
870 0
871 0
872 0
873 0
874 0
875 0
876 0
877 0
878 0
879 0
880 0
881 0
882 0
883 0
884 0
885 0
886 0
887 0
888 0
889 0
890 0
891 0
892 0
893 0
894 0
895 0
896 0
897 0
898 0
899 0
900 0
901 0.00105194
902 0.00302433
903 0.000394477
};
\addlegendentry{JavaScript}
\end{axis}
\end{tikzpicture}
\par}
\caption{Cumulative distribution function (CDF) for the performance of data flow graph extraction, for the two languages. For any number of seconds between 0 and 900 (15 minutes), we depict the ratio of the packages that have finished (or timed out) by then.}
\label{fig:perf_analysis}
